\documentclass[preprint]{aastex62}
\graphicspath{{./}{figures/}}

\received{July 1, 2019}
\submitjournal{AJ}
\usepackage[fleqn]{amsmath} 
\usepackage{amssymb}	
\usepackage{graphicx}	
\usepackage{array}
\usepackage{booktabs}
\usepackage{threeparttable}
\def\degree{${}^{\circ}$}
\shorttitle{Thermal Characteristics of 2008 EV5}
\shortauthors{Jiang H.X. et al.}

\begin{document}
\title{Revisiting Advanced Thermal Physical Model: New Perspectives on Thermophysical Characteristics of (341843) 2008 EV5 from Four-bands WISE Data with Sunlight-Reflection Model}

\correspondingauthor{Jianghui Ji}
\email{jijh@pmo.ac.cn}

\author{Haoxuan Jiang}

\affil{CAS Key Laboratory of Planetary Sciences, Purple Mountain Observatory, Chinese Academy of Sciences, Nanjing 210008, China\\}
\affil{University of Science and Technology of China, Hefei 230026, China\\}

\author{Liangliang Yu}
\affil{State Key Laboratory of Lunar and Planetary Science, Macau University of Science and Technology, Macau, China\\}

\author{Jianghui Ji}
\affil{CAS Key Laboratory of Planetary Sciences, Purple Mountain Observatory, Chinese Academy of Sciences, Nanjing 210008, China\\}
\affil{CAS Center for Excellence in Comparative Planetology, Hefei 230026, China\\}
\nocollaboration



\begin{abstract}

In this paper, we investigate thermophysical characteristics of near-Earth asteroid (341843) 2008 EV5, based on our improved Advanced Thermal Physical Model (ATPM) by considering the contribution of sunlight-reflection for rough surface, along with four wavebands  observations from Wide-field Infrared Survey Explorer (WISE) and the radar-derived shape model.  Here we derive that 2008 EV5 has a relatively low thermal inertia of $\Gamma=110 _{-12}^{+40}\rm~J m^{-2} s^{-1/2} K^{-1}$ but a high roughness fraction. The geometric albedo and effective diameter are then constrained to be $p_v=0.095_{-0.003}^{+0.016}$ and $D_{\rm eff}=431_{-33}^{+6}\rm~m$, respectively. The low thermal inertia indicates that 2008 EV5 may have undergone sufficient space weathering over secular evolution. The high roughness may have resemblances to the appearances of Bennu and Ryugu recently observed by spacecrafts, where a great number of boulders are widely distributed on the asteroid's surface. Moreover, we numerically perform 1000 backward simulations of 2008 EV5's cloned orbits within 1 Myr to explore its origin, and present a probability of $\sim6.1\%$ that the asteroid originates from the main belt.
Finally, we estimate that the mean grain size of the surface ranges from 0.58 to 1.3 mm, and infer that it is unlikely to find water ice on most area of 2008 EV5, but there may exist water ice on high-latitudes near polar region.

\end{abstract}



\section{Introduction} \label{sec:intro}

Over the last decades, space missions to solar system small bodies, such as near-Earth asteroids and comets, have captured the interest of scientists. For these missions, multiple factors should be taken into account when selecting the target small bodies, such as whether a delta-v parameter is low enough to make it more accessible for a rendezvous mission, what composition the asteroid's surface may be made up of, and what types of asteroids may contain important clues to understand the origin and evolution of the early solar system \citep{perna2017}. Asteroid (341843) 2008 EV5 (hereafter EV5) is taxonomically classified as C-type asteroid \citep{reddy2012a}, and is recognized as a potentially hazardous object, which was discovered by the Mount Lemmon Survey (which was a part of the Catalina Sky Survey) in 2008. Because of a small orbital eccentricity of 0.0835 and a low inclination of 7.37\degree, as well as a low delta-v for rendezvous \citep{benner2010}, EV5 was ever proposed to be a potential target for ESA's Asteroid mission MarcoPolo-R and the baseline target for NASA's Asteroid Redirect Mission (ARM). In addition, this asteroid is considered to be the backup target for the future asteroid sample return mission of China National Space Administration (CNSA). Therefore, the size, shape, and surface characteristics of EV5 should be extensively investigated prior to the mission.

Arecibo Goldstone planetary radars and the Very Long Base Line Array (VLBA) observed EV5 in 2008 during its close approach to Earth at a distance of 8.4 lunar distance (approximately 0.022 au) \citep{busch2011}. Using the radar measurements, a high-resolution shape model of EV5 was derived and the diameter for this top-shaped asteroid was given to be $\rm 400\pm50 \ m$ \citep{busch2011}. From the delay-Doppler images, two possible pole orientations were solved, giving J2000 ecliptic longitudes and latitudes of (0\degree, 84\degree) and (180\degree,-84\degree) $ \pm 10$\degree. Moreover, two geological features can be clearly seen from the radar images --- the equatorial ridge which may be induced by YORP acceleration \citep{busch2011,vok2015} and a big concavity with a size roughly one third of the diameter of EV5, which may be caused by an impact event \citep{busch2011}. \citet{reddy2012a} observed EV5 in the visible ($0.30\sim0.92\rm~\mu m$) and near-infrared ($0.75\sim2.5\rm~\mu m$) wavelengths with Palomar Mountain 200-inch telescope and SpeX instrument on the NASA Infrared Telescope Facility (IRTF), respectively, and determined its surface composition. They showed that the spectrum of EV5 shows a weak $0.48\rm~\mu m$ spin forbidden of Fe$^{3+}$. This sort of spectral feature is well consistent with CI carbonaceous chondrites similar to Orgueil. Furthermore, \citet{reddy2012a} presented the albedo of $0.05\sim0.2$ for EV5 based on a modified Standard Thermal Model \citep{spe1989,reddy2012b}, whereas \citet{busch2011} calculated the albedo of $0.12 \pm 0.04$ for the asteroid. These physical parameters provide a good understanding of the evolution for EV5. However, the surface characteristics should be made clear, which plays a vital role in a small body sample return mission.

The surface characteristics of an asteroid, such as the thermal inertia, geometric albedo and roughness, can be revealed by using thermophysical models to fit the mid-infrared observations of the asteroid \citep{roz2011,Delbo2015}. Among these parameters, thermal inertia, defined by $\Gamma=\sqrt{\kappa\rho c}$ (where $\kappa$ is the thermal conductivity, $\rho$ represents the density and c is the thermal capacity), dominates the surface temperature environment of small bodies. This parameter represents the resistance of temperature variations in the surface and subsurface of an asteroid. Moreover, thermal inertia also plays a significant role in Yarkovsky and YORP effects \citep{roz2012}. In the previous study, \citet{Ali2014} showed that EV5 appears to be smooth on its surface with an averaged thermal inertia to be $\sim450\pm60\rm~J m^{-2} s^{-1/2} K^{-1}$  based on the classical thermophysical model (TPM) \citep{lag1996} with W2$\sim$W4 band WISE data.

However, recent close-up observations of small-size asteroids, like (101955) Bennu by NASA's asteroid sample return mission Origins, Spectral Interpretation, Resource Identification and Security-Regolith Explorer (OSIRIS-REx) \citep{Lauretta2019, DellaGiustina2019} and (162173) Ryugu by Hayabusa2 mission \citep{michi2019,Watanabe2019,Sugita2019}, unveil that the surface of these two asteroids bears a high roughness but low thermal inertia, due to the existence of highly space weathered boulders that are widely distributed on the asteroids. In comparison, the shape of EV5 is very similar to the appearances of Bennu or Ryugu. Thus, this raises a key question whether EV5 had experienced similar evolution like Bennu or Ryugu, thereby producing the surface of a high roughness with low thermal inertia. The investigations indicate that low thermal inertia and high roughness behaves similar influence in the radiometry procedure, suggesting that the degeneracy of thermal inertia and roughness fraction may not be well removed by the classical thermophysical models on some occasions. Further analysis is required to shed light on this issue, which enables us to have a better understanding of the thermophysical procedure for small bodies.

The primary objective of this work is to investigate EV5's thermal characteristics and dynamical history from several perspectives.  Firstly, we adopt an alternative technique to screen the data set from WISE archive and all four wavebands (W1 $\sim$ W4) of WISE data . Secondly, we employ our improved Advanced Thermophysical Model \citep{roz2011,yull2014} to derive a new set of thermal parameters for EV5, in which the multiple scattering, self-heating and thermal-infrared beaming effects are taken into account. Besides, in order to remove the effect of reflected sunlight on the results, we propose a new method to evaluate the percentage of reflected part that contains in different wavebands although the reflected part only have marginal effects for $\rm W2 \sim W4$ bands. Based on our results, we show that EV5 bears a high roughness surface with a relatively low thermal inertia to be $\Gamma=110 _{-12}^{+40}\rm~J m^{-2} s^{-1/2} K^{-1}$. The high roughness fraction of EV5 is consistent with the rough surface of Bennu and Ryugu \citep{Lauretta2019, DellaGiustina2019,Watanabe2019,Sugita2019}. Usually the values of thermal inertia for NEAs are relatively large compared to main-belt asteroids \citep{Delbo2015}, such low thermal inertia of EV5 indicates that it may have undergone secular evolution from the main belt region. Therefore, we further study the dynamical evolution of EV5 by performing 1000 backward simulations of its cloned orbits over a timescale of 1 Myr, the numerical results show a probability of $6.1\%$ that EV5 may originate from the main belt region which provides substantial evidence for our radiometric results for this asteroid. Finally, we estimate the mean grain size, and explore the possibility of the existence of water ice on the surface of EV5.

\section{OBSERVATION} \label{sec:observation}
In this work we adopt thermal data of EV5 observed by WISE space telescope, which mapped the entire sky in infrared wavelengths centered at 3.4, 4.6, 12.0 and 22.0 $\rm \mu m$ noted as W1, W2, W3 and W4, respectively \citep{wright2010}. Currently, WISE data are available from WISE archive (http://irsa.ipac.caltech.edu/applications/wise/), which allows us to explore the study. In the procedure, we first achieve the data from the WISE All-Sky Known Solar System Object Possible Association List with a search cone radius of 10" to ensure that the observation target well matches our explored asteroid. Next, we remove those of epochs that cannot provide magnitudes and uncertainties for all four wave bands, to ensure our selected data precise enough for further study. Subsequently, we then convert the magnitudes into fluxes for EV5, following the method described by \citet{wright2010}, where we use the color correction factors of 1.3917, 1.1124, 0.8791 and 0.9865 for $\rm W1\sim W4$ bands, respectively, for this asteroid.

Since the observed fluxes are proportional to the cross-section area of the asteroid in the observation direction, and EV5 is nearly spherical with rotational variations of cross-section area no more than $8\%$, it should thus be expected that the light curves of this asteroid should not have amplitude larger than $30\%$, if considering a maximum observation uncertainties of $\sim20\%$. Therefore we further carry out selections of the observation by considering the amplitude of the WISE thermal light curves. The thermal light curves are derived for each band by correcting the observed flux at various epochs into one rotation period at a reference epoch via \citep{yull2017}
\begin{equation}
\label{fcor}
\begin{split}
 F_{\rm i\_cor} = F_{\rm i}\cdot\left( \frac{r_{\rm helio\_i}}{r_{\rm helio\_0}} \right)^2 \cdot \left( \frac{r_{\rm obs\_i}}{r_{\rm obs\_0}} \right)^2,
\end{split}
\end{equation}
where $F_{\rm i\_cor}$ is the corrected flux, $F_{\rm i}$ is the observed flux at epoch $i$, $r_{\rm helio\_i}$ and $r_{\rm helio\_0}$ are the heliocentric distance of epoch $i$ and the reference epoch, respectively, and $r_{\rm obs}$ represents the distance between the asteroid and the space telescope. Here we use the first epoch in our data set as the reference epoch. Then we remove those data that exceeded $\pm30\%$ of the corrected fluxes' error-weighted  mean value. Finally, we obtained 204 data points that contain all four wavebands with respect to 51 observational epochs, thus producing four thermal light curves at four wavebands. EV5 was monitored by WISE from Jan 25 until March 7, 2010. However, in our screened data set, the observation epochs spread from Jan 25 19:27 to Feb 23 17:22, 2010. The solar phase angle ranges from 71.48 \degree \ to 73.2 \degree.

\begin{figure}
\centering
	\includegraphics[scale=0.3]{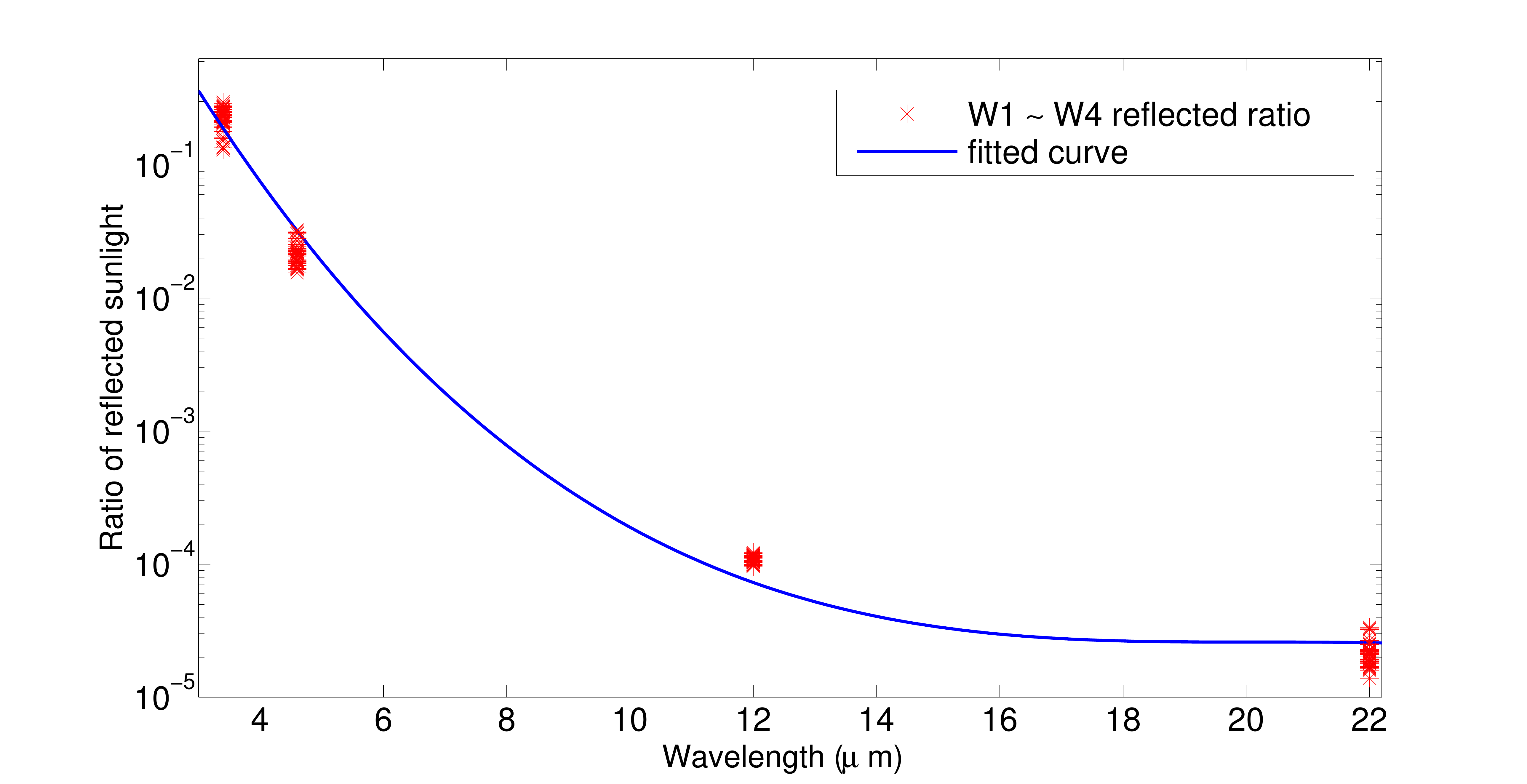}
	\centering
    \caption{The ratio of reflected sunlight to the entire flux in W1-W4 band observation data versus wavelength, 204 data were noted out as red asterisk. Note that most of observations of W1 band contain reflected sunlight in a ratio of about 20 percent but no more than 4\% for W2 band, as to W3 and W4 bands, the reflected sunlight simply contributes to a very small portion of the entire flux. The blue fitted curve was obtained by Polynominal Fitting to express the descending trend of the ratio of reflected sunlight.}
    \label{sunratio}
\end{figure}

It is noteworthy that the WISE data in W1 band contains a significant fraction of reflected sunlight, approximately $20\% \sim 30\%$ of entire flux. As a comparison, the reflected sunlight contributes about $2\% \sim 3\%$ in W2 band whereas it can be negligible in W3 and W4 band, as shown in Fig. \ref{sunratio}, where a sunlight reflection model is applied to calculate the reflected sunlight. The reflection model will be described in detail in the next section.

\section{Radiometric Model}
\subsection{Advanced Thermal Physical Model}
To determine EV5's thermophysical characteristics, we utilize our independently developed codes of the Advanced Thermal Physical Model (ATPM), which has been successfully applied in our previous investigations \citep{yull2017}.
ATPM treats an asteroid as a polyhedron made up by $N$ triangular facets, and can be used to calculate the surface temperature of each triangular facet at each observation epoch by solving one dimensional heat conduction equation under consideration of rough surface boundary. The surface roughness is modeled by a fractional coverage of hemispherical craters to represent the rough surface on each facet while each crater is composed of a certain number $M$ of smaller triangular sub-facets. The shadowing effect, scattering of sunlight and self-heating are taken into consideration in such craters. The roughness fraction is symbolized by $f_{\rm r}$, and thus $1-f_{\rm r}$ represents the flat smooth surface on the asteroid. Both $1-f_{\rm r}$ fraction of the shape facet and sub-facets in the cater can be treated as smooth Lambertian surface. Then the non-linear surface boundary condition can be written as
\begin{equation}
\begin{split}
&(1-A_{\rm B})\ (v\psi F_{\rm Sun}+ F_{\rm scat}) + (1-A_{\rm th})\ F_{\rm rad} +k\ \left( \frac{\rm d \emph{T}}{\rm d \emph{x}}\right)_{\rm x=0}  -\epsilon \sigma  T^{\rm 4}_{\rm x=0} = 0
\end{split}
\end{equation}
where $\kappa$ is the thermal conductivity, $\sigma$ is the Stefan-Boltzmann constant,  $\varepsilon$ is the emissivity which was assumed to be 0.9, $A_{\rm B}$ is the Bond albedo, $\psi$ is the cosine of the solar altitude, x is the depth below the surface, and $v$ represents whether the facet can be seen from the direction of the Sun, it mainly depends on the rotation pole orientation and asteroid's rotation phase. $ F_{\rm Sun}=1367/r_{\rm helio}^2\ \rm W\cdot m^2$ is the incident solar flux to the asteroid at a heliocentric distance of $r_{\rm helio}$ while $F_{\rm scat}$ and $F_{\rm rad}$ are the total scattered and thermal re-emitted flux from other facets respectively, here $r_{\rm helio}$ is in unit of au.

With the temperature distribution, the theoretical thermal emission of each facet can be computed by using the Planck function
\begin{equation}
\begin{split}
B(\lambda,T) = \frac{2 h c^2}{\lambda ^5} \frac{1}{exp(\frac{hc}{\lambda k T})-1},
\end{split}
\end{equation}
where $T$ represents the temperature of the facet. And then the observable thermal emission of the entire body will be
\begin{equation}
\begin{split}
F_{\rm th}(\lambda)=(1-f_{\rm r})\sum_{i}^{N}\epsilon\pi B(\lambda, T_i)S_if_i+f_{\rm r}\sum_{i}^{N}\sum_{j}^{M}\epsilon\pi B(\lambda, T_{ij})S_{ij}f_{ij}
\end{split}
\label{thflux}
\end{equation}
where $\epsilon$ is the monochromatic emissivity at wavelength $\lambda$, $i$ represents the facet $i$ of the shape model, while $ij$ means the sub-facet $j$ in the crater on facet $i$, and hence $S_{i}$ and $S_{ij}$ stand for their areas respectively; $f_i$ and $f_{ij}$ are the their view factors to the telescope. The so-called view factor is defined as
\begin{equation}
f_i=v_i\frac{\vec{n}_{i}\cdot\vec{n}_{\rm obs}}{\pi\Delta^{2}},~
\label{vf}
\end{equation}
where $\vec{n}_{i}$ represents the normal vector of the facet, $\vec{n}_{\rm obs}$ is the unit vector pointing to the observer, $\Delta$ is the observation distance and $v_i$ indicates visible fraction of facet $i$ to the observer.

\subsection{Sunlight-Reflection model of rough surface}
The observed flux is composed by the thermal flux component $ F_{\rm th}$ and the reflected sunlight component $F_{\rm ref}$. As mentioned above, for EV5, the observed W1 band flux contains a significant part of reflected sunlight of $\sim 20\%$. Compared with former studies, in this work we utilize the W1 data and  the reflected sunlight should be corrected. \citet{Ali2013} introduced one technique for calculating the reflected sunlight, which was applied to the near-Earth asteroid thermal model (NEATM) \citep{harris1998}. Based on the method developed by \citet{Ali2013}, \citet{roz2018} re-evaluated the reflected sunlight contained in WISE W2 band while using ATPM to determine the thermal properties of several near Earth asteroids. Besides, \citet{Myh2018} presented a new derivation of simple asteroid thermal model using the NEOWISE data, and the Kirchhoff's law was taken into account when infrared observations contain substantial reflected sunlight. Here we proposed a new method to determine the ratio of reflected sunlight contained in full 4 wavebands although it simply covers a tiny percentage in W2 $\sim$ W4 bands.

On the basis of the aforementioned roughness model, the reflected sunlight can be calculated in a similar way as the thermal emission. So when treating the facet $i$ of the shape model and the sub-facet $j$ in the crater on facet $i$ both to be smooth Lambertian surface, the reflected sunlight from them can be expressed via Lambertian reflection as
\begin{equation}
F_{\rm ref\_i}(\lambda) = B(\lambda,5778)\frac{R_{\rm sun}^2}{r_{\rm helio}^2}\cdot \pi\cdot A_{\rm B}\cdot \psi_i \cdot S_i\cdot f_i,
\end{equation}
\begin{equation}
F_{\rm ref\_ij}(\lambda) = B(\lambda,5778)\frac{R_{\rm sun}^2}{r_{\rm helio}^2}\cdot \pi\cdot A_{\rm B}\cdot\psi_{ij}\cdot S_{ij} \cdot f_{ij},
\end{equation}
where $\psi_i$ and $\psi_{ij}$ are the cosine values of the solar altitudes, $f_i$ and $f_{\rm ij}$ are the view factors defined as Equation (\ref{vf}), $R_{\rm sun}$ is the radius of the Sun, $r_{\rm helio}$ nearly equals to the heliocentric distance of the asteroid, and $B(\lambda,5778)$ is the Planck function under the average temperature (5778 K) of the Sun. Then the total reflected sunlight that can be observed by the telescope is
\begin{equation}
\begin{split}
\label{refl}
F_{\rm ref}(\lambda) =(1-f_{\rm r})\sum_{i}^{N}F_{\rm ref\_i}(\lambda)+f_{\rm r}\sum_{i}^{N} \sum_{j}^{M}F_{\rm ref\_ij}(\lambda).
\end{split}
\end{equation}

Fig. \ref{sunratio} shows the ratio of reflected sunlight contained in each point of observational data. As we can see, the reflected sunlight contained in W3 and W4 bands appears to be very small of only $\sim 10^{-4}$ and $10^{-5}$, whereas it covers a few percent of W2 band. As mentioned previously, the WISE data for W1 band contributes $\sim 20\%$ to the total observed flux, which should  be considered in the model.

\subsection{Fitting procedure}
The radar-derived shape model of EV5 released by \citet{busch2011} contains 3996 facets and 2000 vertexes. Considering the size of the asteroid and the computational efficiency, we further simplified the shape model, which is composed of 1996 facets and 1000 vertexes. The rotation period for EV5 is adopted to be $\rm 3.725\pm0.001 h$ \citep{galad2009}. And the retrograde pole orientation of (180\degree, -84\degree)$\pm10$\degree is also applied. In fact, using the delay-Doppler images, \citet{busch2011} presented a prograde pole orientation of (180\degree, -84\degree)$\pm10$\degree.  By tracking EV5's radar speckle pattern, \citet{busch2010} determined that EV5 rotates in a retrograde mode. Therefore we here do not consider the prograde pole orientation, but use the same pole orientation as \citet{Ali2014} adopted.

Thermal inertia which dominates the surface temperature of the asteroid, and roughness fraction which causes the thermal infrared beaming effect, are two free parameters of this model to fit observation data. In order to find out the best-fitting thermal inertia, the search range of thermal inertia values were set to be  $\Gamma=0 \sim 700$ $ \rm J m^{-2} s^{-1/2} K^{-1}$, and the step size of $\Gamma$  was set to be $\rm 50\ J m^{-2} s^{-1/2} K^{-1}$. For each thermal inertia value, the temperature distribution can be obtained and thus the theoretical thermal emission can be calculated via Equation (\ref{thflux}).
As known, the diameter of an asteroid $D_{\rm eff}$ is correlated with its geometric albedo via \citep{Fowler1992}
\begin{equation}
\begin{split}
\label{deffpv}
D_{\rm eff} = \frac{1329\times10^{-H_{\rm v}/5}}{\sqrt{p_{\rm v}}}.
\end{split}
\end{equation}
On the other hand, an asteroid's effective Bond albedo $A_{\rm EB}$ is correlated to the Bond albedo $A_{\rm B}$ and roughness fraction $f_{\rm r}$ via \citep{wolters2011}
\begin{equation}
\begin{split}
\label{aeff_ab}
A_{\rm EB} = f_{\rm r}\frac{A_{\rm B}}{2-A_{\rm B}}+(1-f_{\rm r})A_{\rm B}.
\end{split}
\end{equation}
Furthermore, the geometric albedo $p_{\rm v}$ can be expressed by
\begin{equation}
\begin{split}
\label{pvaeff}
p_{\rm v} = \frac{A_{\rm EB}}{q_{\rm ph}},
\end{split}
\end{equation}
in which $q_{\rm ph}$ represents the phase integral that was defined by
\begin{equation}
\begin{split}
\label{qphg}
q_{\rm ph}=0.290+0.684G,
\end{split}
\end{equation}
where $G$ is the slope parameter in the $H$, $G$ magnitude system of \citet{bowell1989}. We adopted the value of $H=20.0$ and $G=0.15$ respectively from MPC. Hence, for an input $f_{\rm r}$ and $p_{\rm v}$ we can obtain a unique Bond albedo $A_{\rm B}$ and effective diameter $D_{\rm eff}$.
\begin{table}
	\centering
	\caption{Physical parameters adopted in this work for ATPM.}
	\label{table1}
	\begin{tabular}{lccr} 
		\hline
		Parameter & Value & Reference \\
		\hline
		Number of vertices & 1000   & \citet{busch2011} \\
		Number of facets   & 1996   & \citet{busch2011} \\
		Spin orientation   &  (180\degree, -84\degree) & \citet{busch2011} \\
        rotation period    & 3.725 h  &\citet{galad2009}\\
        Absolute magnitude & 20.0    &MPC\\
        Slope parameter    & 0.15   &MPC\\
        Emissivity        &0.9      &Assumed\\
		\hline
	\end{tabular}
\end{table}

Thus we actually have three parameters to be used in our fitting procedure, geometric albedo $p_{\rm v}$, thermal inertia $\Gamma$ and roughness fraction $f_{\rm r}$. The other input parameters employed are listed in Table ~\ref{table1}. Next, if taking the reflected sunlight Equation~\ref{refl} into consideration, the total flux measured by the observer can be written as
$F_{\rm model}=F_{\rm th} + F_{\rm ref}$
, which can be compared with the observational mid-infrared fluxes from WISE. In order to reappraise the fitting degree of the modeled flux, the reduced $\chi ^2_{\rm r}$ defined by \citet{press2007}
\begin{equation}
\label{chi2r}
\begin{split}
\chi_{\rm r}^2 = \frac{1}{n-3}\sum_{i=1}^{n}\left[\frac{F_{\rm model}(p_{\rm v}, \Gamma, f_{\rm r}, \lambda_{\rm i})-F_{\rm obs}(\lambda_{\rm i})}{\sigma_{\rm \lambda_i}} \right],
\end{split}
\end{equation}
was adopted, where \emph{n} is the number of observation data point, $\sigma_{\rm \lambda_i}$ is the observational flux uncertainty.

\section{Radiometric Results}
\subsection{Thermal inertia and roughness fraction}
\begin{figure}
\centering
	\includegraphics[scale=0.4]{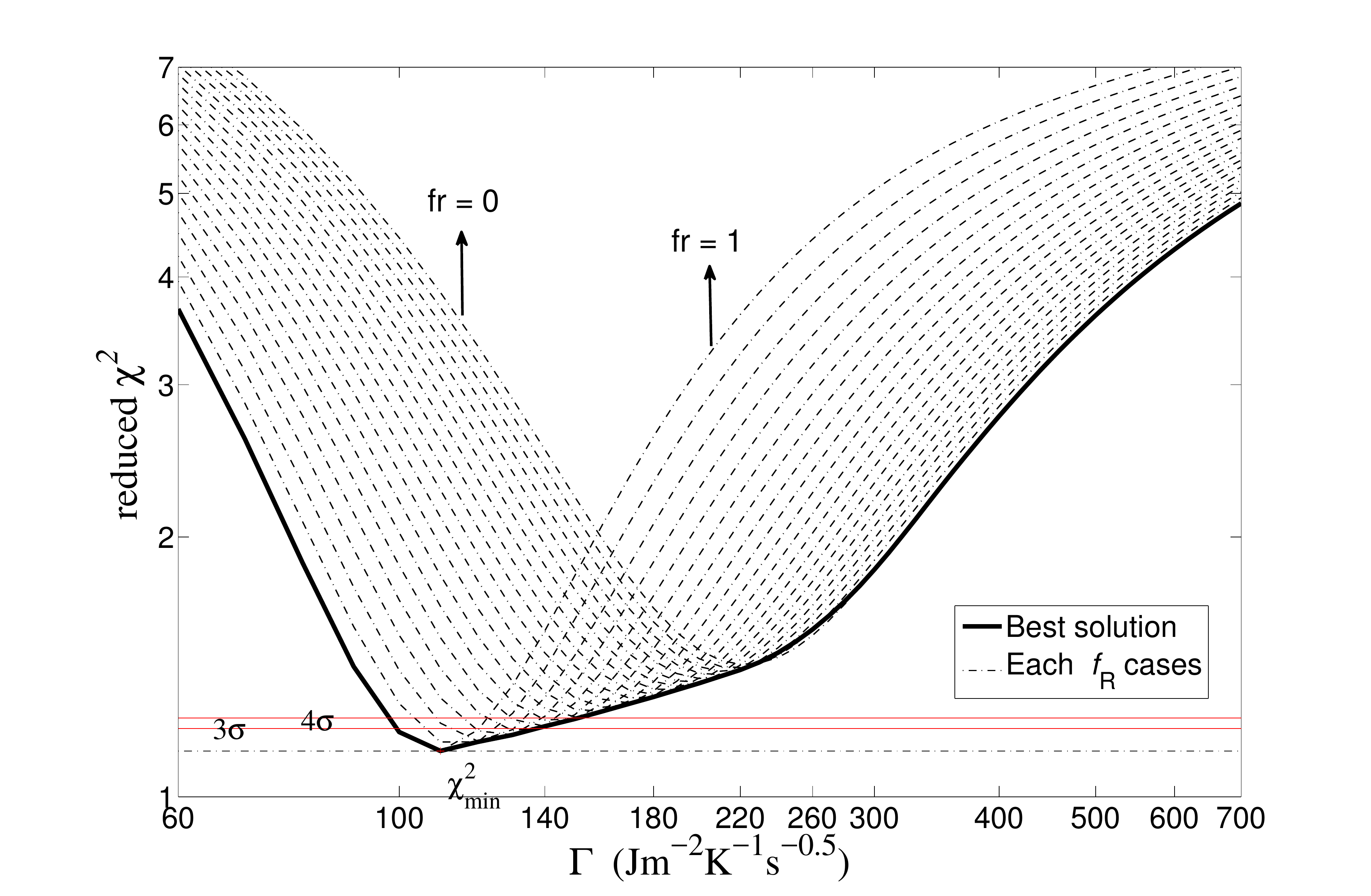}
    \caption{$\Gamma-\chi^2$ profile fit to the observations for EV5. Each dashed curve represents
    a roughness fraction $f_r$  in the range 0.0-1.0, and the red solid lines indicate the $3~\sigma$ and $4~\sigma$
    range for thermal inertia, respectively.}
    \label{chi2_ga}
\end{figure}

In order to obtain the best-fitting value of thermal inertia, $\Gamma-\chi ^2$ profile is plotted in Fig.~\ref{chi2_ga}. As we can see from Fig.~\ref{chi2_ga}, the minimum $\chi ^2= 1.13$ is related to a thermal inertia of 110 $\rm J m^{-2} s^{-1/2} K^{-1}$ with a high roughness fraction. The $3~\sigma$ and $4~\sigma$ ranges are noted as red dashed lines. The lower dashed line labeled $3\sigma$ corresponds to the $\chi^2$ value of $\chi^2_{min} + 14.2/(N-3) = 1.2006$ where $N = 204$ is the number of observations, whereas the upper dashed line labeled $4\sigma$ is associated with the $\chi^2_{min} + 21.0/(N-3) = 1.2345$ \citep{press2007}. Considering the $4~\sigma$ confidence level, we can constrain thermal inertia to be $ 110 _{-12}^{+40}$ $\rm J m^{-2} s^{-1/2} K^{-1}$, whereas the $3~\sigma$ confidence level gives a value of $\rm 110 _{-10}^{+30}\  J m^{-2} s^{-1/2} K^{-1}$. Because of a large number of observations, we use the $4~\sigma$ principle to constrain the uncertainties of our results.

\begin{figure}
\includegraphics[scale=0.4]{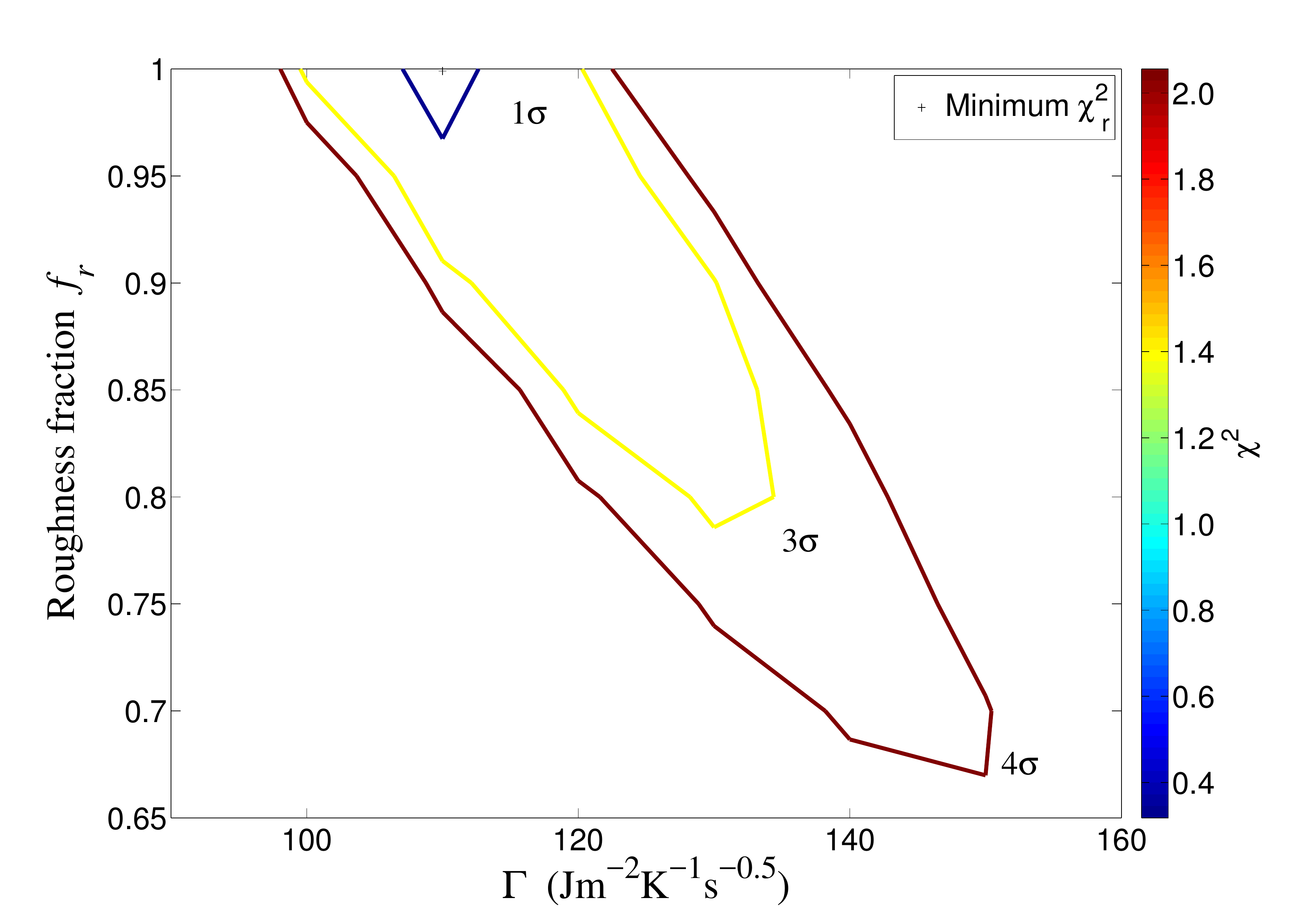}
\centering
 \caption{The $f_{\rm r}-\Gamma$ contour for EV5, where $1~\sigma$, $3~\sigma$ and $4~\sigma$ range are marked in different colors. The minimum $\chi^2$ locates with respect to $\Gamma=110$ $\rm J m^{-2} s^{-1/2} K^{-1} $ with a high roughness fraction $f_{\rm r} = 1.0$. For $4~\sigma$ confidence level, the lower limit of $f_{\rm r}$ is constrained to be 0.68 with respect to the thermal inertia.}

\label{cgcontour}
\end{figure}

\begin{figure}
\centering
	\includegraphics[scale=0.35]{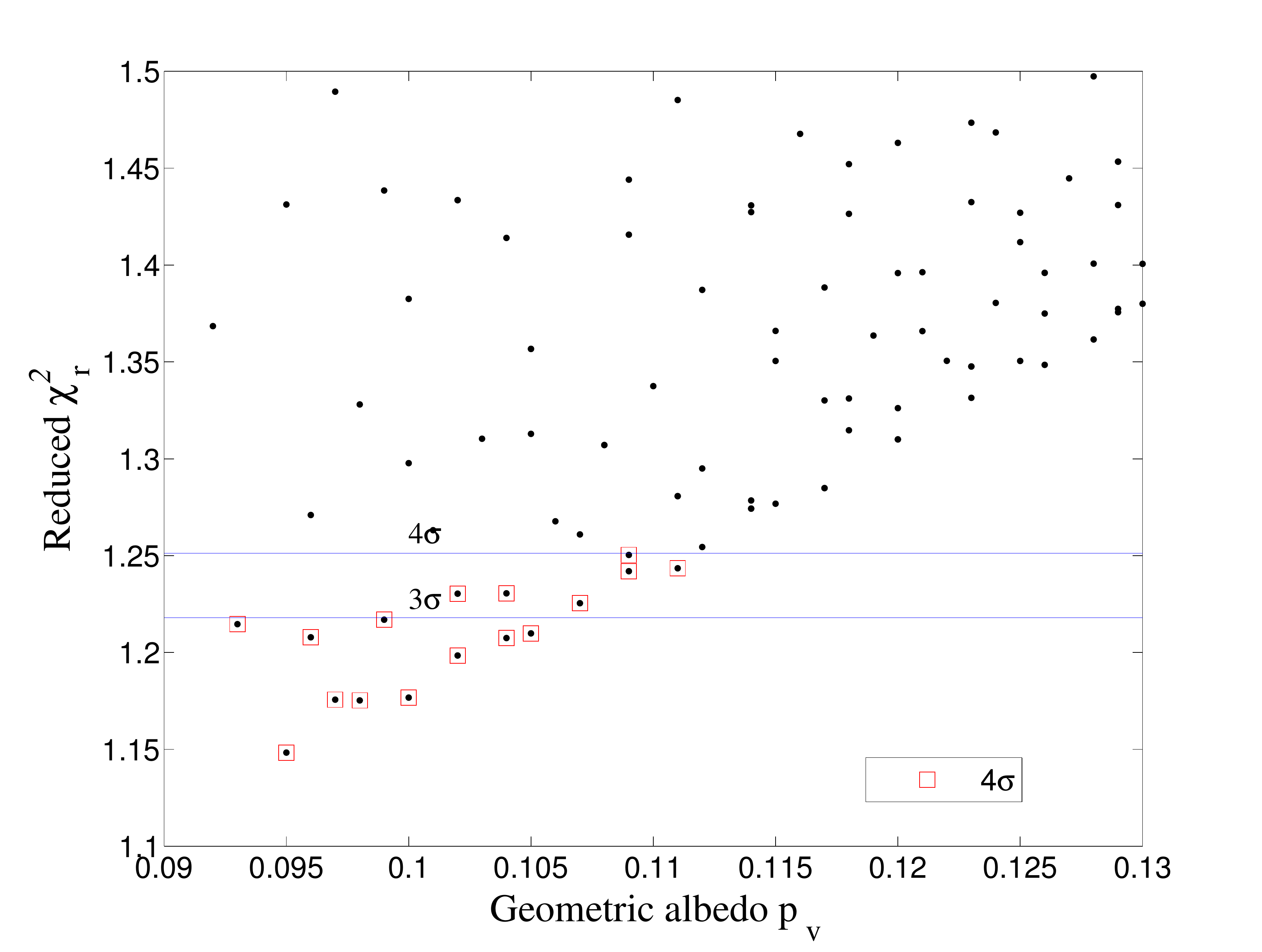}
	\centering
    \caption{The $p_{\rm v}-\chi^2$ distribution for EV5, the minimum $\chi^2$ is correlated to the geometric albedo of 0.095, where the $3~\sigma$ and $4~\sigma$ ranges are plotted as blue lines, and the geometric albedo within $4~\sigma$ range is noted as red squares.}
    \label{pvchi2}
\end{figure}

To further examine the fitting results of thermal inertia, Fig. \ref{cgcontour} shows the contour image of $\chi_{\rm r}^2 (\Gamma, f_{\rm r})$. The blue profile labeled $1\sigma$ indicates the deviation of $\chi^2$ value of $3.52/(N-3)$ from the minimum $\chi^2$ value. And the yellow and red curves are directly concerned with the $3~\sigma$ and $4~\sigma$ confidence level, respectively. As can be noted, the minimum $\chi_{\rm r}^2$ is located at a high roughness fraction and low thermal inertia value of $110$ $\rm J m^{-2} s^{-1/2} K^{-1}$. Moreover, the $\chi^2$ fitting gives the best-fitting geometric albedo of $p_{\rm v} = 0.095 _{-0.003}^{+0.016} $ , using Eq. \ref{deffpv}, where the correlated effective diameter is thus $D_{\rm eff} = 431_{-33}^{+6}\ \rm m$. Furthermore, Fig. \ref{pvchi2} shows the $p_v \sim \chi^2_{\rm r}$ profile for each roughness fraction and thermal inertia derived in the range of $4~\sigma$ level, and the albedo under the $4~\sigma$ range are noted as red squares. Our results are summarized in Table \ref{results}.


\begin{table}
	\centering
	\caption{Thermal parameters derived from ATPM.}
	\label{results}
	\begin{tabular}{lccr} 
		\hline
        \specialrule{0em}{2pt}{2pt}
		Parameter & Value  \\
		\hline
        \specialrule{0em}{2pt}{2pt}
		Thermal inertia $\Gamma$ $\rm (J m^{-2} s^{-1/2} K^{-1})$ & $110 _{-12}^{+40}$  \\
        \specialrule{0em}{2pt}{2pt}
		Effective diameter $D_{\rm eff}$\ (\rm m)& $431_{-33}^{+6}$ \\
        \specialrule{0em}{2pt}{2pt}
		Geometric Albedo $p_{\rm v} $&$0.095 _{-0.003}^{+0.016} $\\
        \specialrule{0em}{2pt}{2pt}
        roughness $f_{\rm r}$&0.68$\sim$1.0\\
		\hline
	\end{tabular}
\end{table}

\begin{figure}
\begin{minipage}{8.5cm}
	\includegraphics[width=\columnwidth,height=5cm]{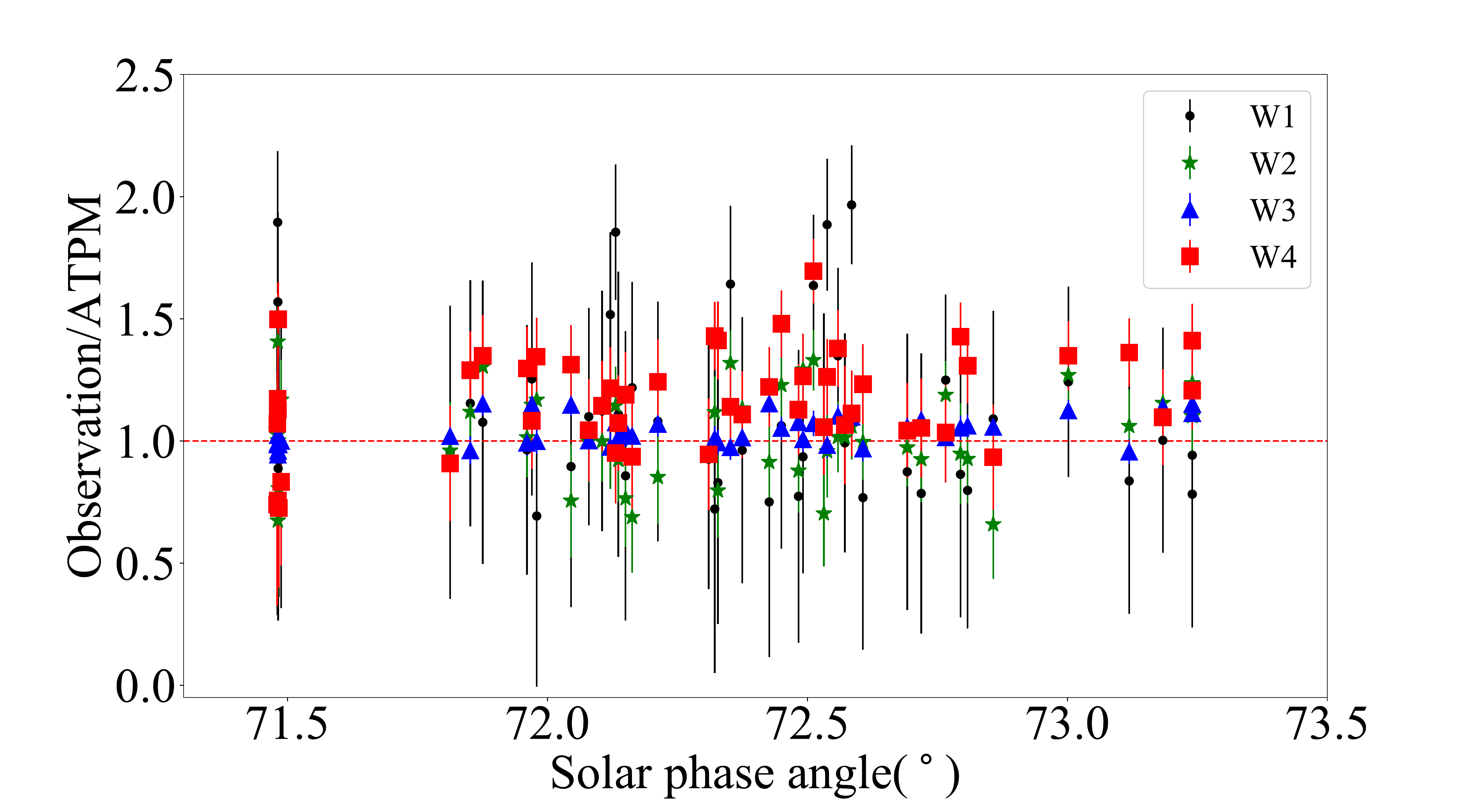}
    \caption{
    The observation/model ratios as a function of solar phase angles for all 4 wave bands.}
    \label{atpm_obs}
\end{minipage}
\begin{minipage}{8.5cm}
	\includegraphics[width=\columnwidth,height=5cm]{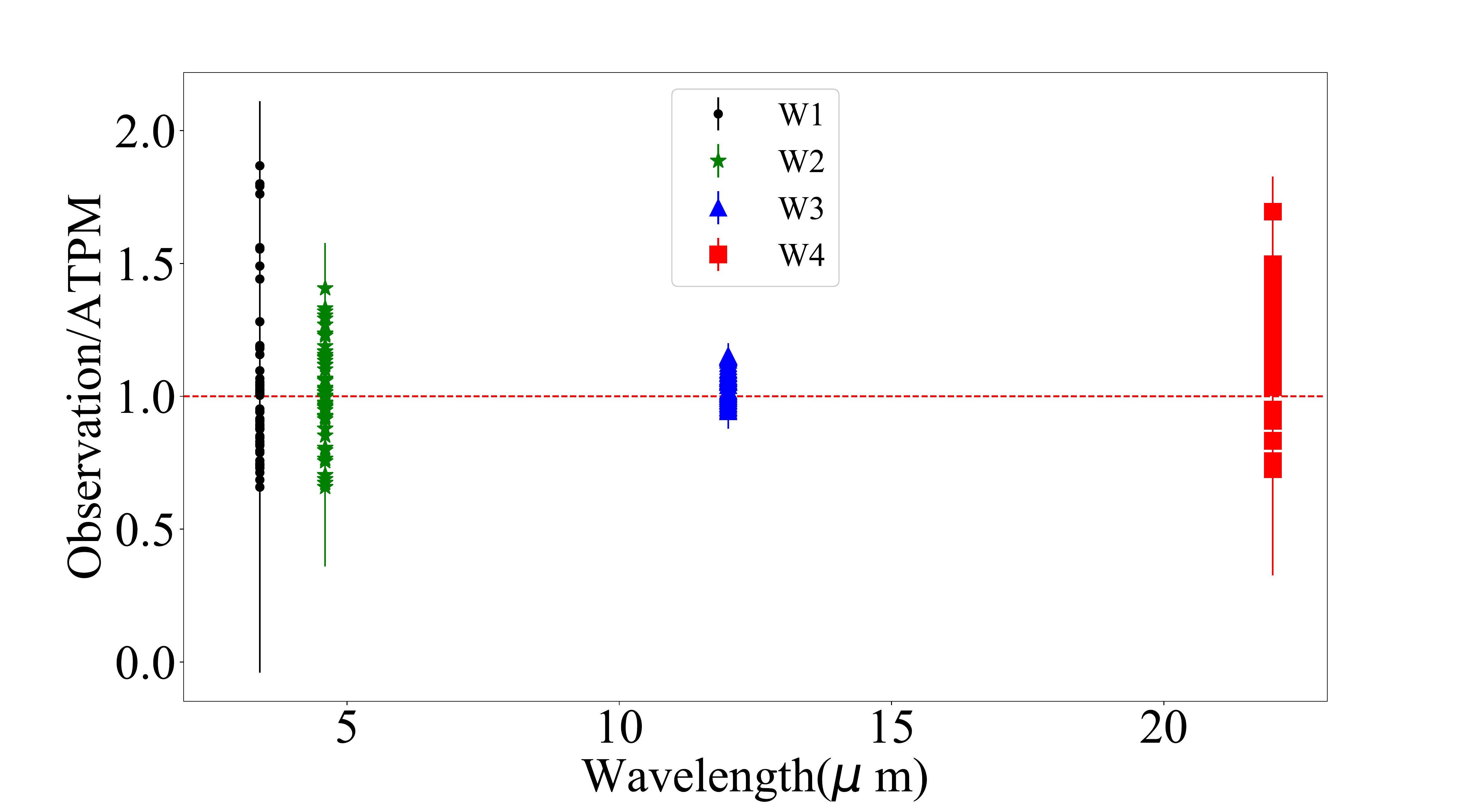}
    \caption{The ratio of theoretical flux to the observation data in all 4 wave bands for different wavelengths.}
    \label{atpm_obs_wavelen}
    \end{minipage}
\end{figure}

To verify that our outcomes are reliable, we compute the ratio of theoretical flux obtained by ATPM and the observational flux, as shown in Fig.~\ref{atpm_obs} and Fig.~\ref{atpm_obs_wavelen}. Note that the "Observation/ATPM" of W1 $\sim$ W4 bands by different colors and the x-axis represents the solar phase angle and wavelengths. As seen from Fig.~\ref{atpm_obs_wavelen}, the ratio of theoretical and observed flux fluctuates at 1 for various observational epochs.


\subsection{Thermal light curve fitting of EV5}

\begin{figure}
\begin{minipage}{9cm}
	\includegraphics[width=\columnwidth]{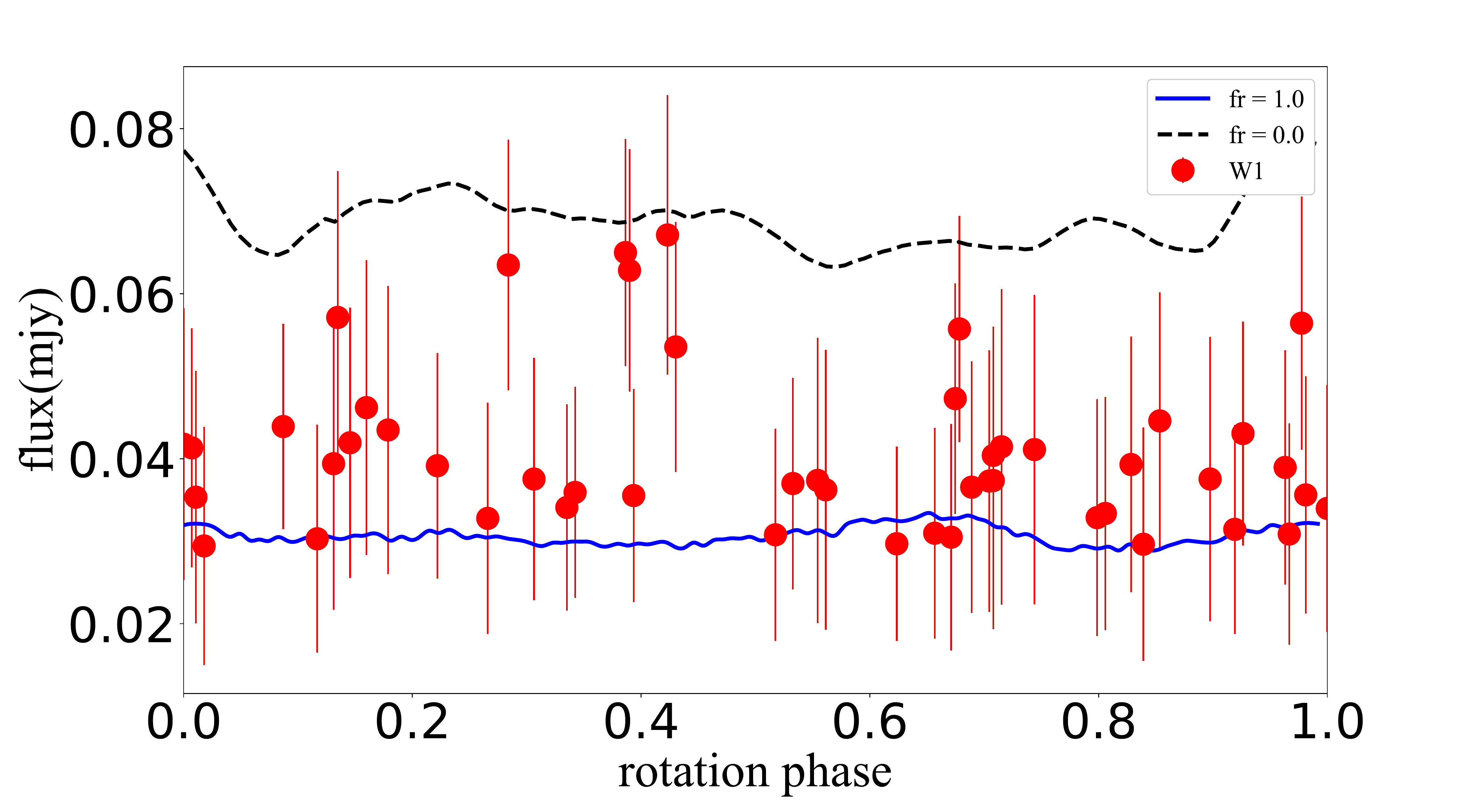}
\end{minipage}
\begin{minipage}{9cm}
	\includegraphics[width=\columnwidth]{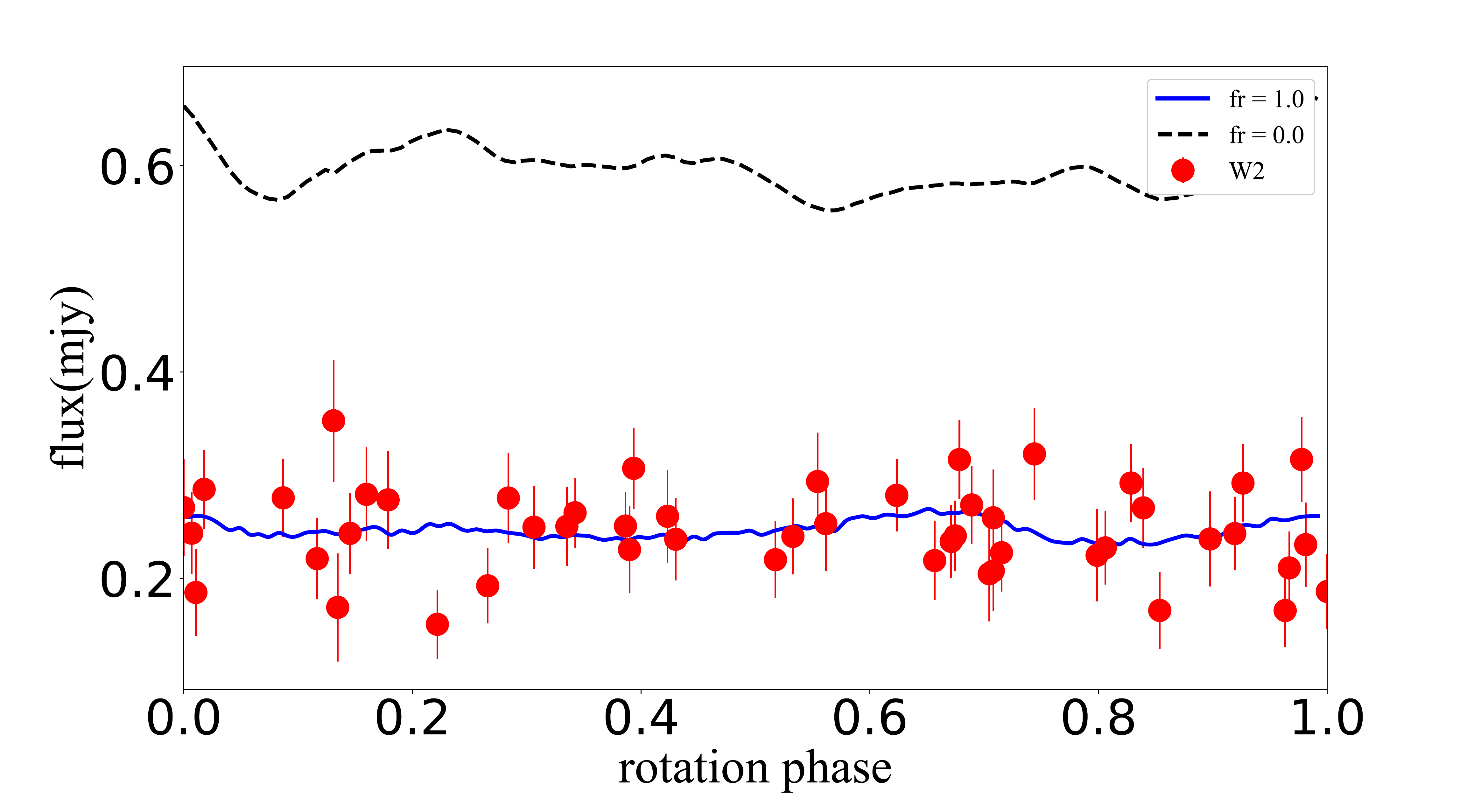}
\end{minipage}
\begin{minipage}{9cm}
	\includegraphics[width=\columnwidth]{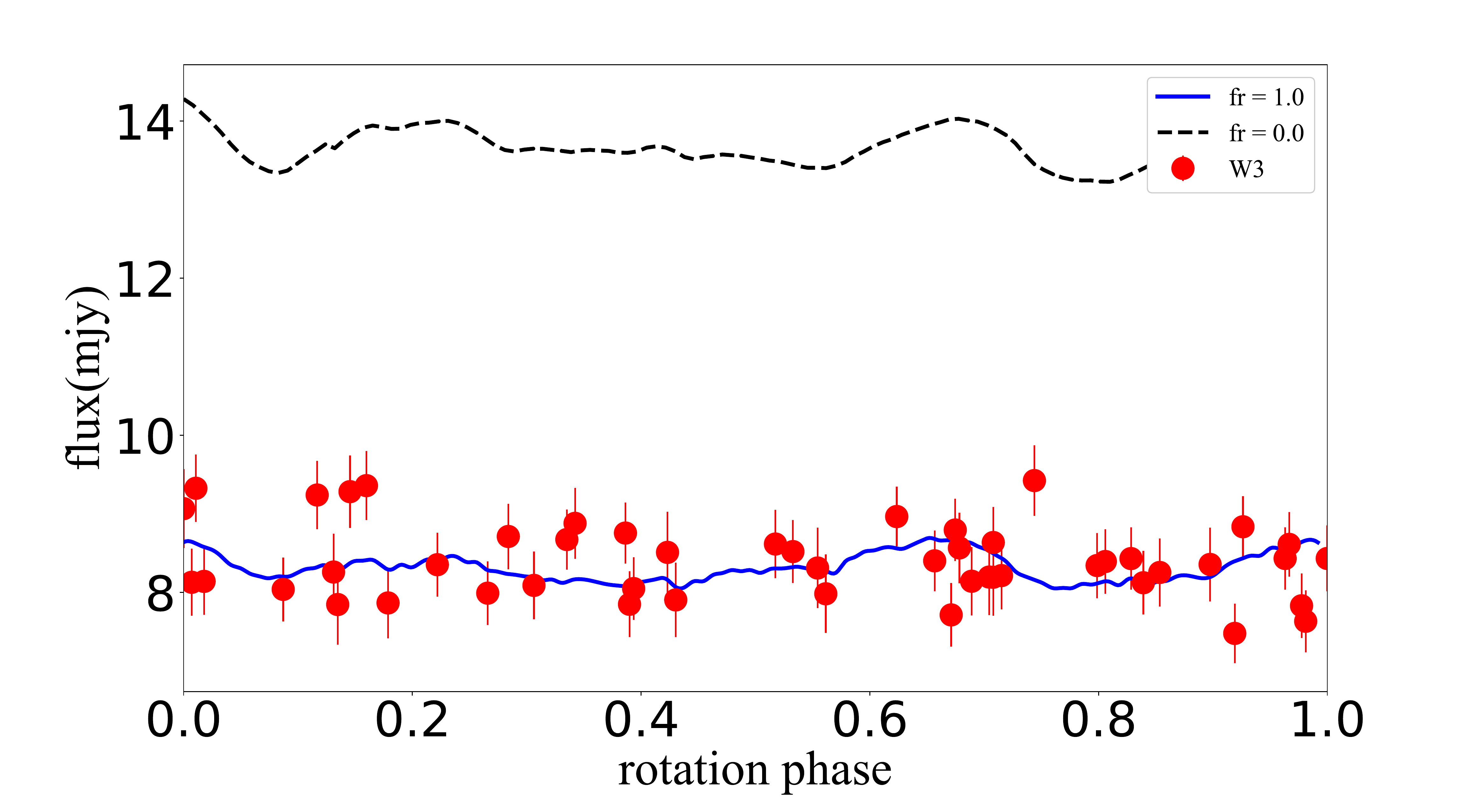}
\end{minipage}
\begin{minipage}{9cm}
	\includegraphics[width=\columnwidth]{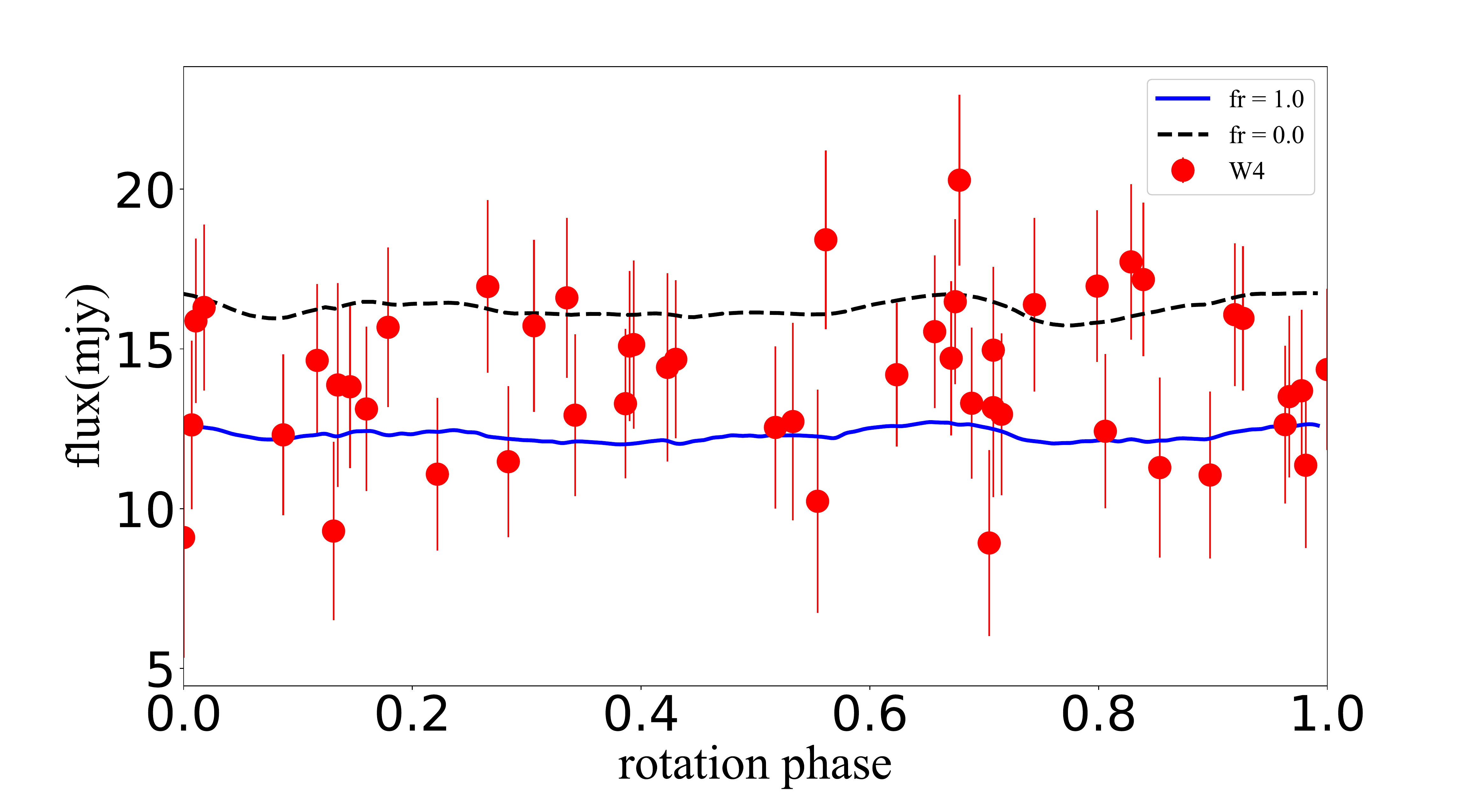}
\end{minipage}
    \caption{Thermal light curves of W1$\sim$W4 bands. The red error bars represent the observations converted into the first epoch in our dataset using Eq.~\ref{fcor}; the blue solid curves refer to the thermal light curves for high roughness fraction while the black solid curves for low roughness cases, both rough and smooth cases are modeled with $\Gamma=110$ $ \rm J\cdot m^{-2}\cdot s^{-1/2} \cdot K^{-1}$. The rough thermal light curves match the observations better than the smooth curves.}
    \label{thermallight}
\end{figure}

Although the WISE data we used in this work covers long consecutive observation epochs, the time interval between two adjacent observation epochs is larger than that of the rotation period of only 3.725 hours. In this case, these data cannot be used to directly produce thermal light curves of the asteroid. Hence we applied the method used in \citet{yull2017} to generate the rotation phase and thermal light curves of EV5. In the study we adopted the extensive radar shape model of EV5 to define the local body-fixed coordinate system. If the telescope viewed the asteroid in a direction of $(\Phi, \theta)$, where $\Phi$ and $\theta$ are local longitude and local latitude respectively, then the rotation phase of one observational epoch can be written as
\begin{equation}
\begin{split}
phase = 1 - \frac{\Phi}{2 \pi}.
\end{split}
\end{equation} 
We assume EV5's rotation phase at epoch 2010-01-25 19:27 to be $zph$, the rotational phase at other epochs can be deduced according to the observation time and geometry. Besides, the observed flux of other epochs were also converted by using Eq.~\ref{fcor}. With the derived rotational phases and the converted data, we can compare with the modeled thermal light curve as shown in Fig.~\ref{thermallight}. The red error bars are observational data of W1 $\sim$ W4 bands and the blue curves are modeled thermal light curves. In Fig.~\ref{thermallight} we can see that the modeled light curves changes periodically as the asteroid rotates, and high roughness results can better fit to the thermal lightcurves as well.

Here we compare our results with those of \citet{Ali2014}. The major difference may be interpreted as follows. First, we use an alternative method to screen the WISE dataset, where the $\rm W1 \sim W4$ bands data are here adopted for fitting. Second, in this work we adopt the modified advanced thermophysical model (ATPM), where the modeling of roughness differs from that of the classical thermophysical model (TPM), thus it is very different for the response of theoretical thermal flux to vary with the input roughness and thermal inertia for ATPM and TPM. Hence, both of aspects can lead to alternative results from the optimization fitting process between two models. As well-known, roughness and thermal inertia show similar influence in the radiometry procedure, only if the degeneracy of thermal inertia and roughness fraction is well removed, the derived radiometric results are reliable. According to our fitting results, either high roughness with low thermal inertia or low roughness with high thermal inertia (close to the result of \citet{Ali2014}) can be well fitted to the observations. But in our model, the case of high roughness can produce a minimum reduced $\chi^2$, we believe that high roughness with low thermal inertia for EV5 can better explain the observations.  Furthermore, as shown in Fig. \ref{thermallight}, the thermal lightcurves of W1, W2 and W3 bands with high roughness show best-fitting outcomes with respect to the observational data, whereas for low roughness fraction, the thermal lightcurves have a poor fitting degree with the observations. While in 22 $\rm \mu m$ of W4, the theoretical flux of both high roughness and low roughness cross the observational data. Hence, we conclude that, the $\rm W1 \sim W3$  data may mostly contribute to the result of $f_{\rm r} = 1.0$. In addition, as Fig. \ref{sunratio} shown, in W1 band, the reflected sunlight $F_{\rm ref}$  contributes to $\sim$ 20\% of the observational flux, whereas in $\rm W2 \sim W4$ bands, the reflected sunlight $F_{\rm ref}$ can be negligible. If we remove $F_{\rm ref}$ from $F_{\rm model}$, we find that the thermal inertia is approximately 100 $ \rm  J m^{-2} s^{-1/2} K^{-1}$ with a geometric albedo $0.095_{-0.003}^{+0.012}$, which is slightly different from those of considering the reflected sunlight. However, in such case the value of $\chi^{2}_{\rm min}$ is larger than that of Fig.~\ref{chi2_ga}.

The weak influence of reflected sunlight here may be further explained. According to Wein��s displacement law ($\lambda_{\rm max} = b/T$, where $b = 2898$  $ \rm \mu m\cdot K$ is the Wein��s displacement constant and $T$ the temperature), if we consider $T = 300 $ K, the spectral radiance of EV5 peaks at about 10 $\rm \mu m$, which means the S/N ratio of W1 and W4 will be smaller than that of W2 and W3. Under such circumstance, the uncertainties of W1 and W4 will be larger than those of W2 and W3, thereby resulting in the low weight of W1 data in the fitting process which in turn reduces the contribution from reflected sunlight. However, for other asteroids with different temperature from EV5, the uncertainties in W1 may be smaller than that of EV5, thus the reflected sunlight could have a significant effect of the fitting result. Moreover, the reflected part here plays an essential role in improving the fitting degree.

\section{Simulations of Dynamical history}
As the thermal inertia we obtained is quite small for a NEA, we infer that this asteroid may be a rubble-pile asteroid, which has already been formed for a long time, thus the rocks or boulders on its surface had experienced a long time of space weathering and thermal fatigue \citep{Delbo2014,Delbo2015}, thereby even showing low thermal inertia but still remaining high roughness. In order to examine the possibility of our inference, we will investigate the dynamical history of EV5 to see whether the asteroid originates from the outer region of the solar system. Here, we simply concentrate on orbital information within a specific time interval before the current epoch, and the evolution time span was chosen to be 1 Myr. In our extended simulations, we use the widely-adopted package MERCURY \citep{Chambers1999} to numerically integrate the orbits of EV5.

Here we adopt the initial orbital elements of nominal orbit and their uncertainties for EV5 from the Minor Planet Center (MPC)(see Table \ref{orbitelement}). Similar to \citet{han2016}, we generated 1000 orbital clones except for the nominal orbit. All the orbital elements in each clone was yielded within the uncertainties given in Table \ref{orbitelement}. The initial epoch was set to be $\rm MJD = 58482.03712$, in which the asteroid was near the perihelion. We adopt the time step to be 7 days in the integrations. In addition, the perturbations from eight major planets and Pluto in solar system were considered in the model (without including the non-gravitational forces, influences from large asteroids such as Ceres, Pallas and Vesta) when we integrate backward 1000 cloned orbits. The ephemerides of these major planets are from JPL DE405 with initial epoch of  MJD = 51000.

\begin{table}
	\centering
	\caption{Orbital elements and uncertainties for EV5 adopted from MPC (MJD = 58482.03712).}
	\label{orbitelement}
    \begin{threeparttable}
	\begin{tabular}{lccr} 
		\hline
		\hline
        \specialrule{0em}{2pt}{2pt}
		Element & Value & Uncertainty \\
		\hline
         \specialrule{0em}{2pt}{2pt}
        \specialrule{0em}{2pt}{2pt}
		semimajor axis (au) & 0.9582193  & $4.0598\times 10^{-8}$\\
        \specialrule{0em}{2pt}{2pt}
		eccentricity & 0.0833817 & $3.8432\times10^{-8}$\\
		\specialrule{0em}{2pt}{2pt}
		inclination (\degree)&7.43739&$1.7923\times 10^{-6}$\\
        \specialrule{0em}{2pt}{2pt}
        argument of perihelion (\degree)&234.85528&$1.1770 \times 10^{-5}$\\
        \specialrule{0em}{2pt}{2pt}
        ascending node (\degree)&93.38202&$1.7171\times10^{-6}$\\
         \specialrule{0em}{2pt}{2pt}
         mean anomaly (\degree) &0&-\\
		\hline
	\end{tabular}
    \begin{tablenotes}
    \footnotesize
    \item[]Note: since our initial position was  perihelion, the value of initial mean anomaly was set to be zero.
    \end{tablenotes}
	\end{threeparttable}
\end{table}

\begin{figure*}
	\includegraphics[scale=0.28]{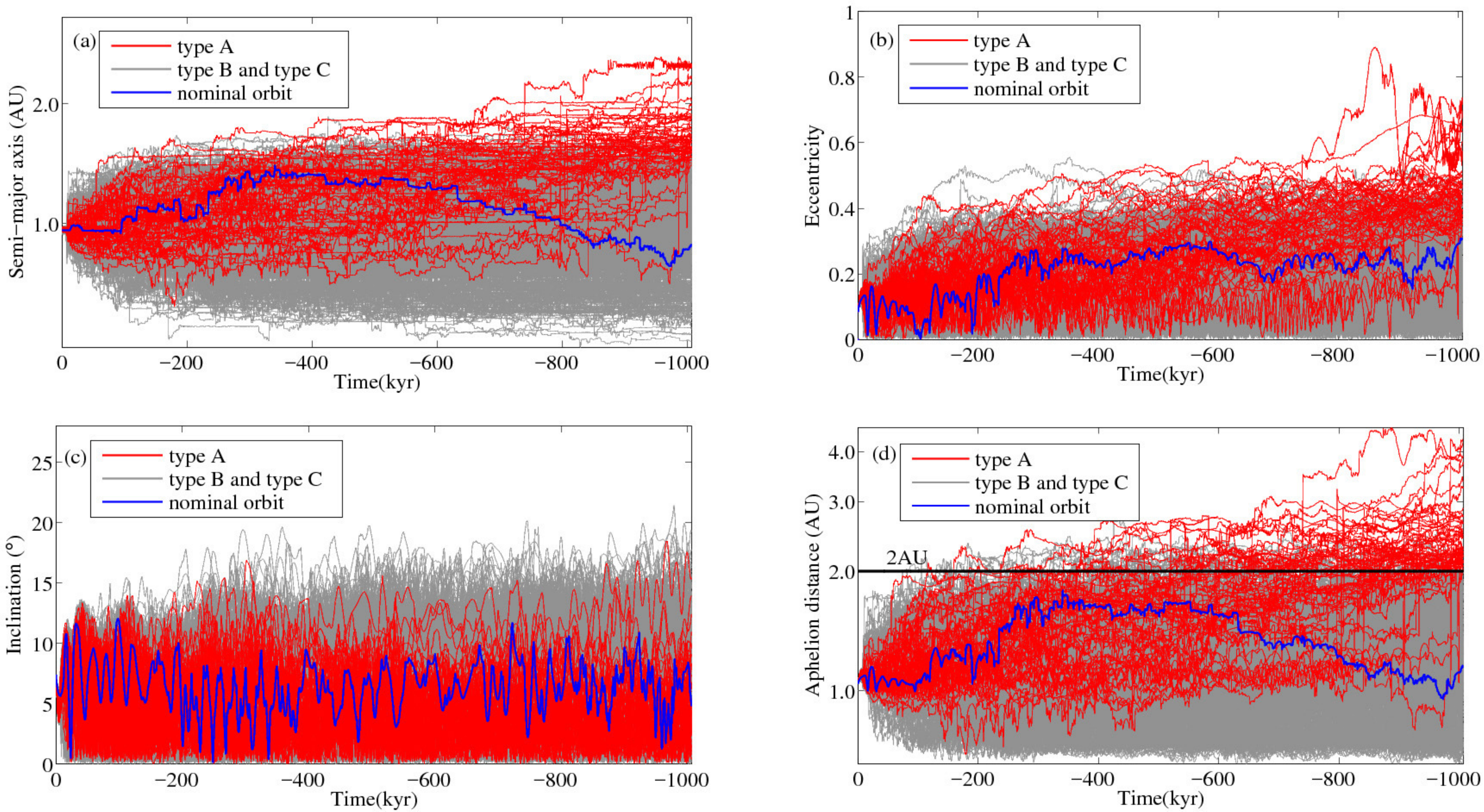}
    \caption{Backward integration results of EV5's clone orbits: (a) the semi-major axis's evolution, (b) the evolution of  eccentricity, (c) the inclination variations, (d) the aphelion distances versus time. The red lines,  gray lines, respectively,  refer to type A orbit,  type B and type C orbit. The blue line shows the evolutions of nominal orbit. The black solid line represents the distance of 2 au.}
    \label{orbitresult}
\end{figure*}

Fig. \ref{orbitresult} shows the backward integration results of EV5's clone orbits within 1 Myr, where Panel (a), (b), (c) and (d), each refer to the trace-back evolution of the semi-major axis, eccentricity, inclination and aphelion distances for  test particles, respectively. As a comparison, the blue curves are relevant to orbital evolution of the nominal orbit for EV5. As shown in Fig. \ref{orbitresult}, we observe that even a tiny difference in initial orbital parameters can eventually give rise to diverse evolution outcomes for these orbits. To obtain the evidence that EV5 originates from the main belt or outer region, we need to find out whether there exists a solution that the orbital aphelion can reach the main belt region (i.e. $a\cdot(1+e)\geq 2.0 \rm au$ , where $a$ is the semi-major axis and $e$  the eccentricity).  Because of a small eccentricity of EV5, the semi-major axis and aphelion distances show some similarities in their evolution profiles.

For convenience, we define three types of test orbits noted as type A, type B and type C, respectively. Type A represents the orbits that can reach the main belt region at the end of 1 Myr in the backward integration, type B stands for the orbits with aphelion distances temporarily larger than 2 au at a certain time, while type C denotes the orbits that cannot reach the main belt region in the simulations. In the following, we will characterize the results of the clone orbits according to three types.

\begin{figure}
	\includegraphics[scale=0.14]{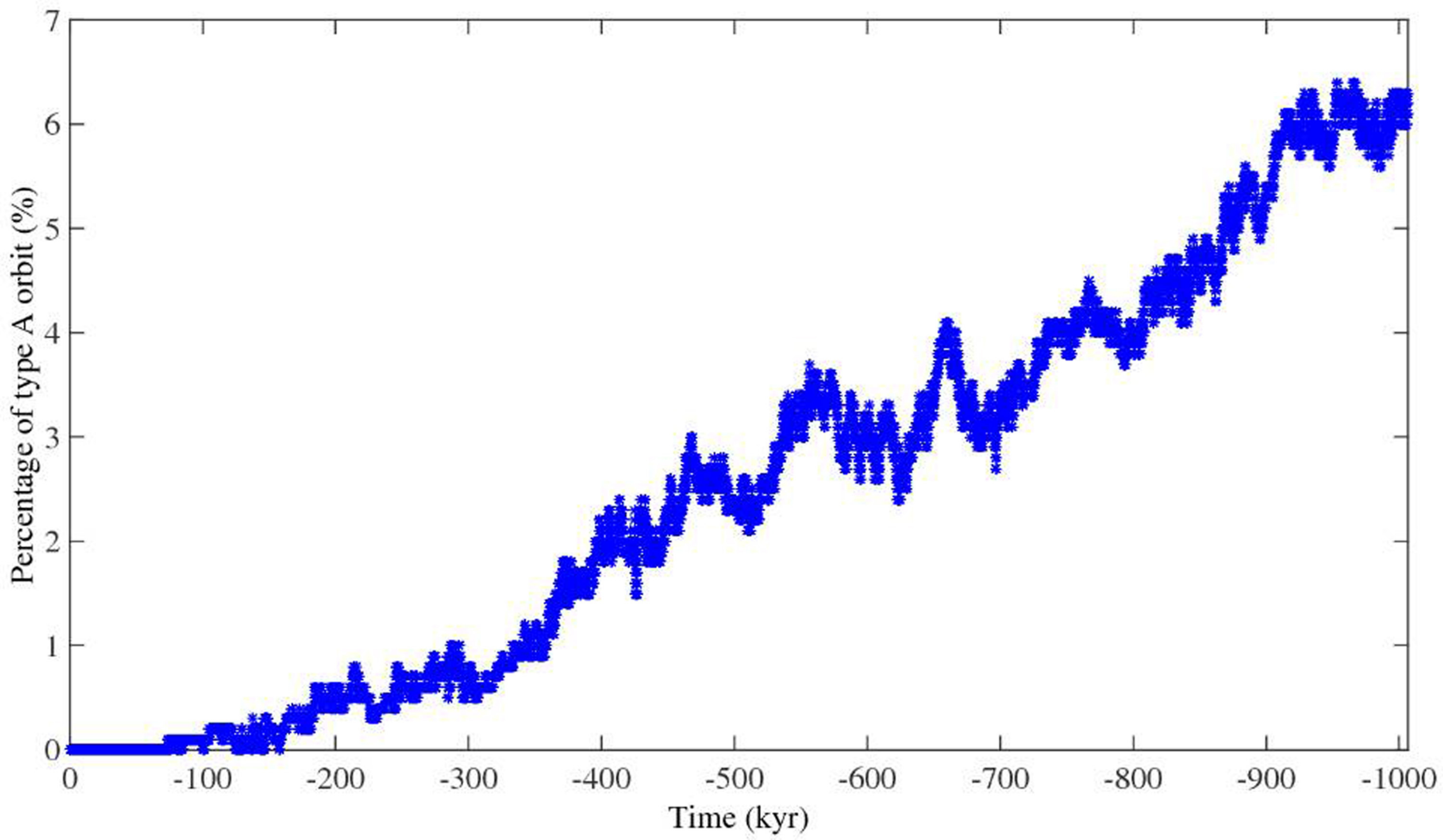}
	\centering
    \caption{Variation of percentage of type A orbit. The fluctuations of the percentage indicate the existence of type B orbit because of the reduction of aphelion distances of test clones during the evolution histories.}
    \label{pertypea}
\end{figure}

At the backward integration of 1 Myr, the aphelion distances of some orbital clones are even larger than 4 au, which were completely different from the nominal orbit. For some orbital clones, their eccentricities can be stirred up to several times of the current value (i.e. about 0.083). By analyzing the simulations results, we have counted the number of type A orbit to be 61, type B orbit of 105, and type C orbit of 834. The simulations indicate that this asteroid has a probability of 6.1 percent from the main belt. Besides, from Fig. \ref{orbitresult}(a), we can see that for some type A orbit, the semi-major axis could be smaller than 0.7 au during the past 1 Myr. According to $q=a\cdot(1-e)$ and combined with the excited eccentricity, the perihelion distance $q$ could be even smaller than 0.3 au. This means the asteroid could be closer to the Sun than Mercury.  Therefore, the large range of heliocentric distance allows the materials on its surface to experience long-term evolution (such as thermal fatigue, dehydration, etc), resulting in a low thermal inertia. As shown in Fig. \ref{pertypea}, we notice that the cumulative number of type A orbit varies with time, giving an increasing trend in the probability.

It is usually postulated that the main-belt asteroids with larger diameters tend to have smaller thermal inertia \citep{Delbo2009}. For example, \citet{yull2017} presented the thermal inertia and effective diameter of (349) Dembowska to be $20_{-7}^{+12}$ $\rm  J m^{-2} s^{-1/2} K^{-1}$ and $155.8_{-6.2}^{+7.5}$ km using ATPM and thermal data from WISE, Subaru, Akari and IRAS. \citet{hor2012} derived the thermal inertia and effective diameter of Trojan asteroid (1173) Anchises to be 25-100 $\rm  J m^{-2} s^{-1/2} K^{-1}$ and $136_{-11}^{+18}$ km . Most of these asteroids have thermal inertia no more than 200 $\rm  J m^{-2} s^{-1/2} K^{-1}$. Our result of thermal inertia for EV5 is consistent with those of \citet{hor2012}. Hence, we infer that EV5 may originate from the main-belt region and was one of the fragments of a larger asteroid. Thus the dynamical history of this asteroid could be longer enough to produce a fine regolith layer with a low thermal inertia.

\section{Predictions}
\subsection{Mean grain size --- dust exist?}
Since EV5 is the potential target for the future sample return mission, we are also interested in the asteroid's surface feature, such as whether there exists dust layers on its surface and how the dust grain sizes behave. In this work, based on the thermal inertia we derived and the method by \citet{gun2013}, we can further estimate the dust grain size of EV5.

First, the relation between thermal conductivity and thermal inertia can be written as
\begin{equation}
\begin{split}
\kappa =\frac{\Gamma^2}{(1-\phi) \rho c},
\end{split}
\label{grain1}
\end{equation}
where $\phi$ is porosity, $\rho$ is the material density and $c$ is the specific heat capacity. The thermal conductivity could also be derived by a theoretical model developed for granular materials in vacuum \citep{gun2012}, which can be given as
\begin{equation}
\begin{split}
\kappa = \kappa _{\rm solid}\left[\frac{9 \pi}{4}\frac{1-\mu^2}{E}\frac{\gamma(T)}{b}\right]^{1/3}\cdot[f_1e^{(f2 (1-\phi))}]\cdot\chi + 8\sigma \epsilon T^3\frac{e_1\phi}{1-\phi}b.
\end{split}
\label{grain2}
\end{equation}

\begin{figure}
\centering
	\includegraphics[scale=0.35]{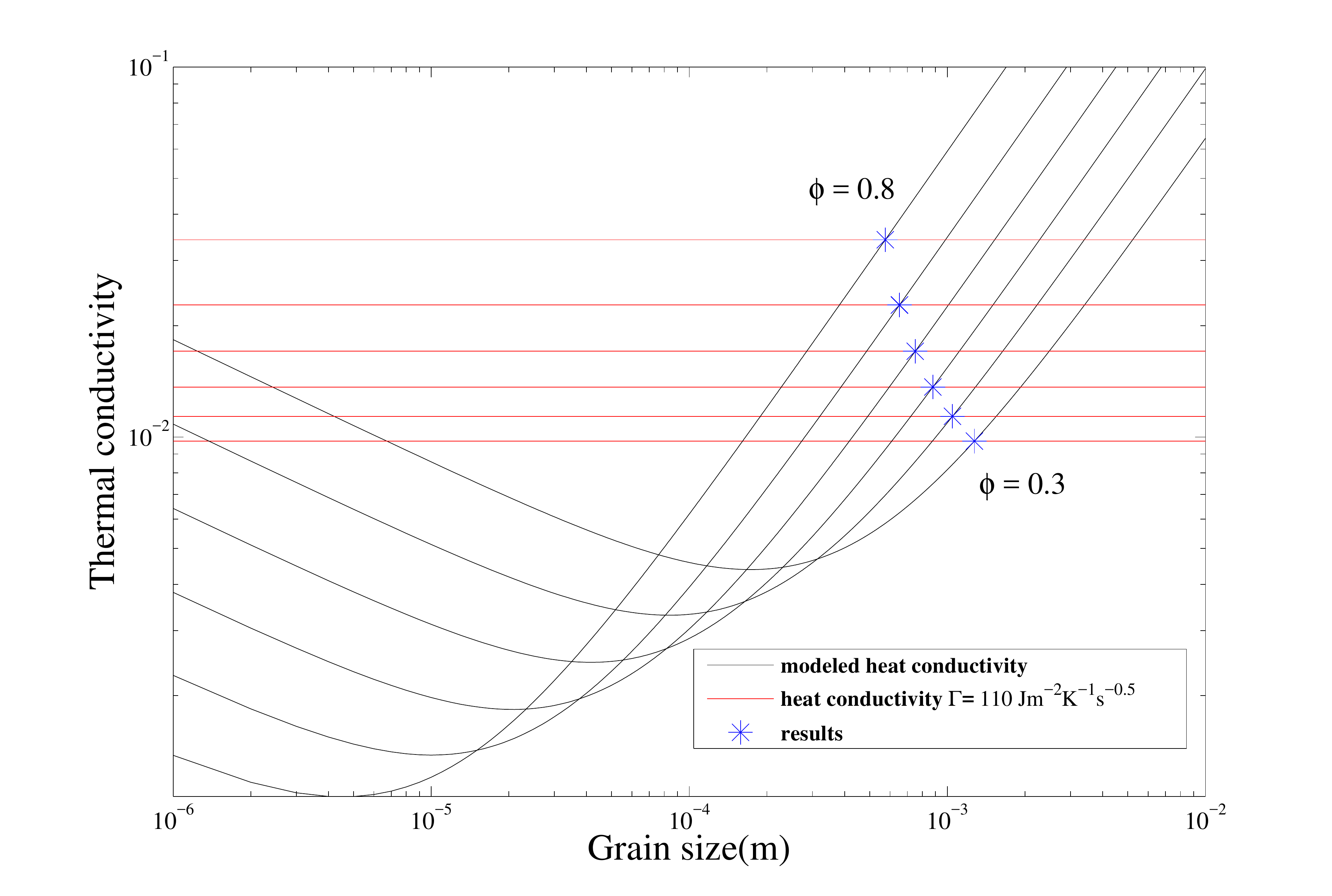}
    \caption{Heat conductivity derived from the thermal inertia (red lines) and by heat conductivity model (black curves), the blue dot represents the result particle sizes.}
    \label{grainsize}
\end{figure}

The quantities we used in Eqs. \ref{grain1} and \ref{grain2} are listed in  Table \ref{grainpa}. $b$ represents the mean grain radius of interest. Fig. \ref{grainsize} shows the $b \sim \kappa$ curve based on the two equations above with the parameters in Table \ref{grainpa}. The results of different value of volume filling factors are noted as blue asterisks. Because EV5's orbit moves close to the Earth's orbit, so we assume the surface temperature to be 300 K, and other parameters are adopted from \citet{gun2013}. We estimate the grain radius to be 0.58 - 1.3 mm based on the thermal inertia value of $\Gamma = 110\rm~J m^{-2} s^{-1/2} K^{-1}$. Our results of the grain sizes differ from those of \citet{Ali2014} in that we use the different assumption of thermal inertia.

This value of grain radius suggests at least the existence of dust aggregates on the surface of EV5. Considering the high roughness, these dust aggregates may be produced from boulders by space weathering or thermal fatigue \citep{Delbo2014}, but being coherent there and thus do not fall down to form a smooth regolith layer across the surface. As a result, thermal inertia of the boulders becomes much lower but the roughness still remains high.

\begin{table}
	\centering
	\caption{Physical parameters used to derive the grain size.}
	\label{grainpa}
    \begin{threeparttable}
	\begin{tabular}{lc|cc} 
		\hline
		\hline
        \specialrule{0em}{2pt}{2pt}
		Parameter & Value  &Parameter &Value \\
		\hline
        \specialrule{0em}{2pt}{2pt}
		$\kappa_{\rm solid}\tnote{}$& $1.19+2.1\times 10^{-3}T \ [\rm Wm^{-1}K^{-1}]$ & $e_1$&$1.34\pm0.01$ \\
        \specialrule{0em}{2pt}{2pt}
		$\mu \tnote{}$ & 0.25&$\epsilon\tnote{}$&$0.9$\\
        \specialrule{0em}{2pt}{2pt}
		E\tnote{}& $\rm 7.8\times 10^{10} [Pa] $&$\rho \tnote{}$&$3110$\\
        \specialrule{0em}{2pt}{2pt}
        $\gamma(T)\tnote{}$ &$6.67\times 10^{-5}T \ [\rm Jm^{-2}]$&$T\tnote{}$&$300 \rm K$\\
         \specialrule{0em}{2pt}{2pt}
		$f_1$&$0.0518\pm0.0345$&$c \tnote{}$&$570$\\
         \specialrule{0em}{2pt}{2pt}
		$f_2$&$5.26\pm0.94$&$\phi$&$0.3\sim 0.8$\\
        \specialrule{0em}{2pt}{2pt}
		$\chi$&$0.41\pm0.02$\\
		\hline
        \specialrule{0em}{2pt}{2pt}
        Notes:
	\end{tabular}    \begin{tablenotes}
    \footnotesize
    \item[] $\kappa_{\rm solid}$: granular heat conductivity of solid material
    \item[] $\mu$: Poisson's ratio
    \item[] E: Young's modulus
    \item[] $\gamma(T)$: specific surface energy of the grain material
    \item[] $f_1,f_2$: empirical constants
    \item[] $\chi$: factor describes the reduction of $\kappa$
    \item[] $e_1$: scaling factor to estimate the mean free path of the photons
    \item[] $\epsilon$: emisivity
     \item[] $\rho$: density of solid material
      \item[] \emph{c} heat capacity of solid material
      \item[] \emph{T} surface temperature
      \item[] $\phi$: porosity
    \end{tablenotes}
\end{threeparttable}
\end{table}

\subsection{Can EV5 have water ice retained?}
If the asteroid comes from the main belt, EV5 may be initially formed as an icy small body due to a C-type object from spectral analysis. The C-type asteroids are mostly believed to be primitive, volatile-rich residuals from the early Solar System \citep{Rivkin2002}.
Therefore, we are very curious about whether it is likely for EV5 to have water ice retained as of today at its present near-Earth orbit. To answer this issue, we assume EV5 to be an icy small body that bears a homogeneous ice-dust mixture interior with a dust mantle covered on top, then apply the "Dust-ice Two-layer Model" of icy small bodies built in \citet{yull2019}. On the basis of previous results, the assumed physical parameters for the dust mantle for EV5 are listed in Table \ref{phpa}.
\begin{table}
 \centering
 \renewcommand\arraystretch{1.3}
 \caption{Assumed physical parameters of EV5's dust mantle.}
 \label{phpa}
 \begin{tabular}{@{}lcc@{}}
 \hline
 \hline
 Property & Value \\
 \hline
 Grain density $\rho_{\rm d}$  & $3110\rm~kgm^{-3}$  &  \\
 Mean Grain radius $b$        &  $1\rm~\rm mm$      &  \\
 Tortuosity $\varsigma$  &  2      &  \\
 Porosity $\phi$          &  0.4      &  \\
 Ice/dust ratio  $\chi_0$ &  0.1$\sim$0.2      &  \\
 \hline
\end{tabular}
\end{table}

\begin{figure}
\begin{minipage}{8.5cm}
\includegraphics[scale=0.58]{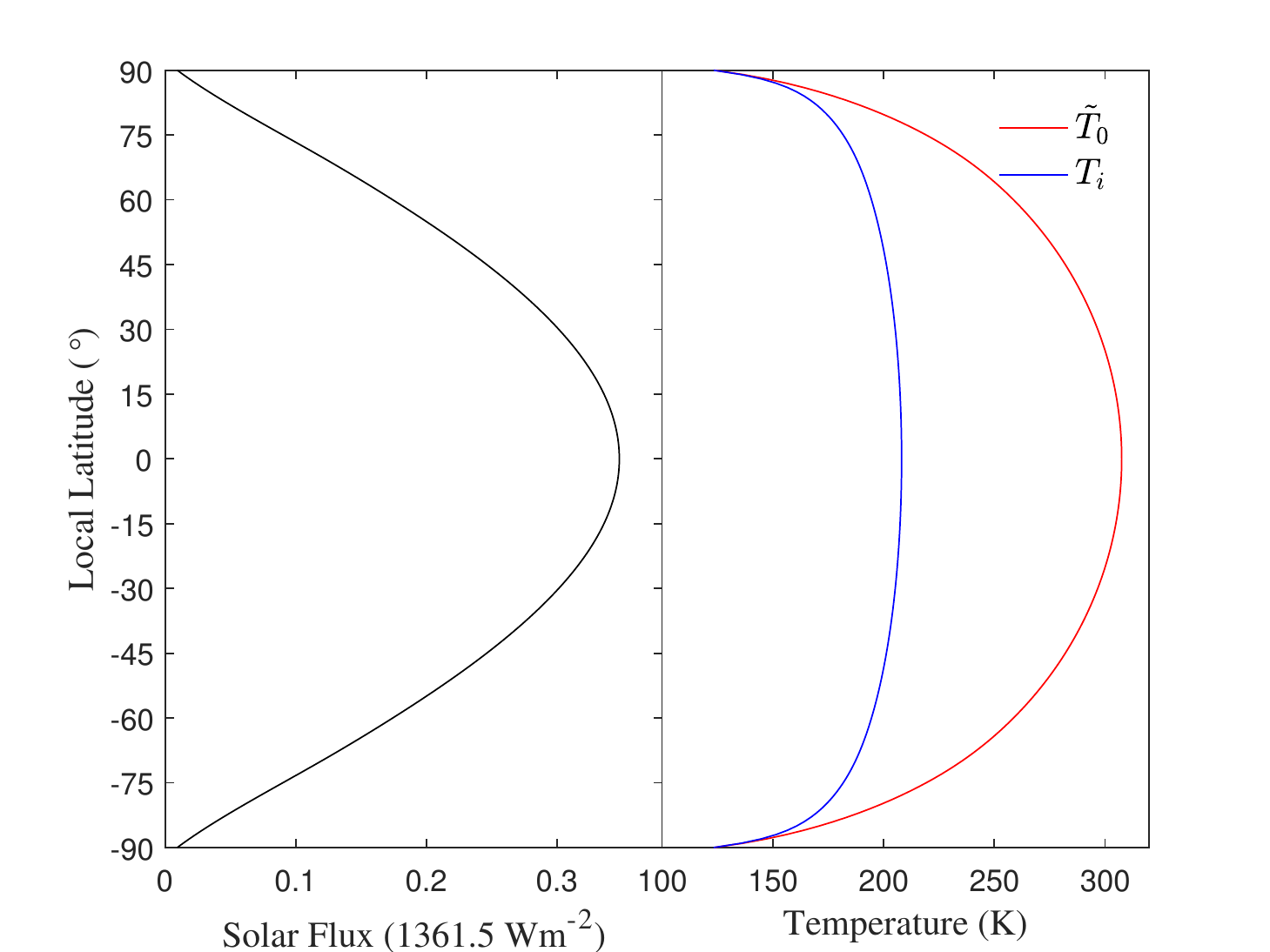}
  \centering
  \caption{Left panel: Annual mean solar insolation as a function of local latitude on EV5.
  Right panel: Equilibrium subsurface temperature $\tilde{T}_0$ and ice front (if ever there was) temperature
  $T_{\rm i}$ , assuming a geometric albedo of $p_{\rm v}=0.09$.
  }\label{AFAT}
\end{minipage}
\begin{minipage}{8.5cm}
\includegraphics[scale=0.58]{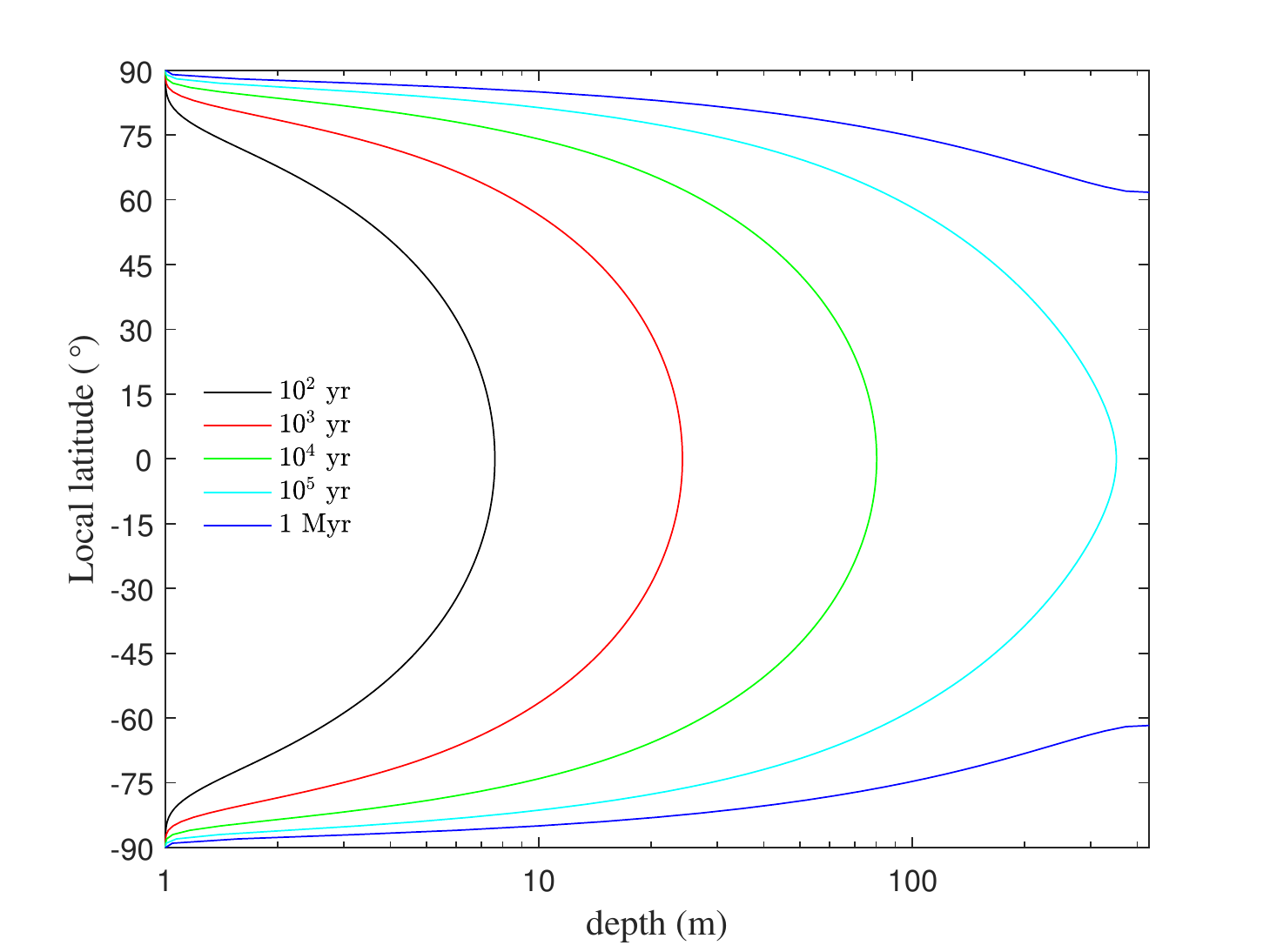}
  \centering
  \caption{Secular inward motion of the ice front.
  }\label{IceFrontHt}
 \end{minipage}
\end{figure}

The rotation axis of EV5 is nearly perpendicular to its orbital plane, which leads to
the polar regions that can receive less solar insolation than the equatorial regions (the left panel in Fig. \ref{AFAT}), and thus show lower temperature. In the right panel of Fig. \ref{AFAT}, the red curve shows the equilibrium subsurface temperature $\tilde{T}_0$ on each local latitude of EV5, whereas the blue profile represents the relevant temperature $T_i$ of ice front (the interface between the dust mantle and the icy interior). The equatorial subsurface temperature is approximately $\tilde{T}_0=308\rm~K$, much higher than the corresponding ice front temperature $T_i=208\rm~K$, while on the polar regions $T_i$ is nearly equal to $\tilde{T}_0$ with a very low value of $\sim124\rm~K$.
Thus the larger temperature gradient in the dust mantle at equator would make the water ice there sublimate and escape more quickly, nevertheless the low temperature environment of polar regions would be good reservoirs for water to retain.

Given the ice sublimation front temperature, we can trace the secular
evolution of the ice front in EV5, as shown in Fig. \ref{IceFrontHt}. We find that the ice at equatorial region cannot be retained for a time scale $\sim10^5$ yr for its present orbit. And for most low-latitude regions ($<$ 60$^\circ$), the ice cannot be maintained longer than 1 Myr. Therefore, if EV5 is transferred from the main belt region to the present orbit within $\sim1\rm~Myr$, it is unlikely to discover water ice on most region of EV5 except the high-latitudes near polar region ($>$ 60$^\circ$).

\section{CONCLUSION AND DISCUSSION}
In this work, we explore thermal characteristics of EV5 using our newly improved Advanced Thermal Physical Model (ATPM) by considering the contribution of reflected sunlight, in combination with the 3-D radar-derived shape model \citep{busch2011} and the 4-bands observations from WISE. We show that EV5 has a thermal inertia of $110 _{-12}^{+40}$ $ \rm  J m^{-2} s^{-1/2} K^{-1}$, a geometric albedo $p_{\rm v} = 0.095 _{-0.003}^{+0.016} $, an effective diameter $D_{\rm eff} = 431_{-33}^{+6}$ m, and a high roughness fraction through the computations. As a comparison, the effective diameter is consistent with that of \citet{busch2011}, but the values of thermal inertia and geometric albedo given in this work are smaller than those of \citet{Ali2014}. The derived thermal inertia indicates the mean grain size of the surface materials ranging from 0.58 to 1.3 mm.

For a small-size NEA, it is generally believed to be a rubble-pile object, where the surface of which should be made up by a wide variety of size of boulders, thereby having an averaged high thermal inertia and high roughness initially. Our findings suggest that EV5 may have a low thermal inertia with a high roughness. The likely interpretation for such low thermal inertia may be induced by the scenario that the surface had suffered an ever-lasting history of space weathering and thermal fatigue \citep{Delbo2014, Delbo2015}, after it had been formed for a certain long time. Under such circumstance, dust aggregates or smaller fragments can be produced from the boulders, thereby making the surface have a lower thermal inertia. From the extensive backward simulations of 1000 clone orbits for EV5 for 1 Myr, we show that there is a likelihood of $6.1\%$ that this asteroid can originate from the main belt or the outer region in our solar system, which is supportive of the low thermal inertia in this work. However, during the secular evolution history, the surface can still remain high roughness (see Fig. \ref{thermallight}), being indicative of that most of the produced dust aggregates should be coherent on the parent boulders and thus did not fall down to form a smooth regolith layer across the surface. \citet{sanch2014} showed that rubble-pile asteroids have nonzero cohesive strength due to small van der Waals forces between constituent grains, and this strength allows some aggregates or smaller fragments remain where they were produced, therefore it would not alter the high roughness of surface. Moreover, such kind of low thermal inertia but high rough surface texture recently happens on Bennu \citep{Lauretta2019,DellaGiustina2019} and Ryugu \citep{ Sugita2019, Watanabe2019} as observed by OSIRIS-REx and Hayabusa2, respectively, which provides strong supports for our inference and results.

The prediction of possibility of finding water ice on high-latitudes near polar region of EV5 is based on the current small axial tilt $\sim10^\circ$ between its rotational axis and orbital axis, but largely depends on the historical axial tilt. So the solution may be diverse if EV5 has a totally different axial tilt in the past. Nevertheless, for current EV5, the polar regions are good reservoirs for water ice to retain due to low temperature. If the rotation axis changes in the future, the water ice can sublimate to release gas and then produce activity. As a matter of fact, recent observations provide direct evidence that several B-type asteroids, e.g., (3200) Phaethon \citep{Devogele2018,yull2019} and  Bennu \citep{Cellino2018} may be active asteroids or dormant comets. Recently, OSIRIS-REx did discover ejection events of dust particles from Bennu's surface (www.asteroidmission.org/?latest-news=nasa-mission-reveals-asteroid-big-surprises;
www.planetary.org/blogs/jason-davis/osiris-rex-update-bennu-ejection.html),
thereby raising the interesting issue - whether Bennu would be a comet, although the mechanism of ejection events is not clear yet.

In conclusion, we speculate EV5 may have a low thermal inertia but high rough surface, being indicative of a long lasting evolution timescale in the main belt. There may exist dust aggregates on the surface and water ice on the polar regions. Hence, EV5 appears to be a suitable target for a sample return mission to retrieve high reward in science in future.



\acknowledgments
We thank the referee for constructive comments and suggestions.
This work is financially supported by the National Natural Science Foundation of China (Grant Nos. 11473073, 11661161013, 11633009), CAS Interdisciplinary Innovation Team, the Strategic Priority Research Program on Space Science, the Chinese Academy of Sciences,
Grant No. XDA15020302) and Foundation of Minor Planets of the Purple Mountain Observatory. This research has made use of the NASA/ IPAC Infrared Science Archive, which is operated by the Jet Propulsion Laboratory, California Institute of Technology, under contract with the National Aeronautics and Space Administration. Research using WISE Release data is eligible for proposals to the NASA ROSES Astrophysics Data Analysis Program.

\end{document}